\documentclass[aps,pra,reprint,superscriptaddress,showpacs]{revtex4-1}
\usepackage[a4paper,left=1.75cm,right=1.75cm,top=2cm,bottom=2cm]{geometry}
\usepackage{amsfonts,amsmath,amssymb,mathrsfs,mathtools}
\newcommand{\maths}[1]{$\smash{#1}$}
\usepackage{xcolor,graphicx}
\usepackage[pdfstartview=FitH,colorlinks=true,linkcolor=blue,citecolor=blue,urlcolor=blue]{hyperref}

\begin{document}

\title{Postquench prethermalization in a disordered quantum fluid of light}

\author{Pierre-\'Elie Larr\'e}
\email{pierre-elie.larre@u-cergy.fr}
\email{pierre-elie.larre@lkb.upmc.fr}
\affiliation{Laboratoire de Physique Th\'eorique et Mod\'elisation, Universit\'e de Cergy-Pontoise, CNRS, 2 Avenue Adolphe-Chauvin, 95302 Cergy-Pontoise CEDEX, France}
\affiliation{Laboratoire Kastler-Brossel, Sorbonne Universit\'e, CNRS, ENS-Universit\'e PSL, Coll\`ege de France, 4 Place Jussieu, 75252 Paris CEDEX 05, France}

\author{Dominique Delande}
\affiliation{Laboratoire Kastler-Brossel, Sorbonne Universit\'e, CNRS, ENS-Universit\'e PSL, Coll\`ege de France, 4 Place Jussieu, 75252 Paris CEDEX 05, France}

\author{Nicolas Cherroret}
\affiliation{Laboratoire Kastler-Brossel, Sorbonne Universit\'e, CNRS, ENS-Universit\'e PSL, Coll\`ege de France, 4 Place Jussieu, 75252 Paris CEDEX 05, France}

\date{\today}

\begin{abstract}
We study the coherence of a disordered and interacting quantum light field after propagation along a nonlinear optical fiber. Disorder is generated by a cross-phase modulation with a randomized auxiliary classical light field, while interactions are induced by self-phase modulation. When penetrating the fiber from free space, the incoming quantum light undergoes a disorder and interaction quench. By calculating the coherence function of the transmitted quantum light, we show that the decoherence induced by the quench spreads in a light-cone fashion in the nonequilibrium many-body quantum system, leaving the latter prethermalize with peculiar features originating from disorder.
\end{abstract}


\maketitle

\section{Introduction}
\label{Sec:Introduction}

Recent avant-garde experiments on cold atomic vapors \cite{Greiner2002, Kinoshita2006, Sadler2006, Hofferberth2007, Cheneau2012, Trotzky2012} attracted wide interest in the thermalization of many-body quantum systems projected away from equilibrium after a quench. In a typical setup, a many-body quantum system is initially prepared in the ground state of a given Hamiltonian and is suddenly forced to evolve according to a time-modified version of this Hamiltonian (quench protocol). Due to particle interactions, the associated energy difference is redistributed among the degrees of freedom of the system and the latter relaxes towards a thermal equilibrium state (thermalization process). Depending on the system, this stationary state is predicted to be described by either a Gibbs or a generalized Gibbs density matrix, the temperature of which is set by the energy injected into the system \cite{Srednicki1994, Rigol2007, Kollath2007, Barthel2008, Rigol2008, Polkovnikov2011}.

If the thermalized regime is in general well understood, the nonequilibrium dynamics leading to this state still puzzles. In particular, are there peculiar stages explored by the system before it fully thermalizes? It was shown that generic many-body quantum systems first relax towards a quasistationary thermal state usually referred to as prethermalized \cite{Berges2004}. In such systems, the actual thermalization emerges only later on, when inelastic scattering becomes nonnegligible. Prethermalization was studied in various condensed-matter systems ranging from quantum Ising chains \cite{Marino2012, VanDenWorm2013, Marcuzzi2013, Marino2014} to Bose-Hubbard gases \cite{Moeckel2008, Eckstein2009, Moeckel2009, Moeckel2010}, Bose-Einstein condensates \cite{Kitagawa2011, Barnett2011, Gring2012, Kuhnert2013, Langen2013}, and Tomonaga-Luttinger liquids \cite{Mitra2013, Buchhold2016}.

If the above descriptions hold for homogeneous many-body quantum systems, they should be considered cautiously when disorder is present. As a matter of fact, due to the complex interplay between interactions, which drive thermalization, and disorder, which induces localization, it is not at all guaranteed that the equilibrium state of a many-body quantum system evolving in a disordered landscape can be described within a standard statistical-mechanics framework \cite{Gornyi2005, Basko2006}. This phenomenon, known as many-body localization, is currently under active investigation \cite{Nandkishore2015, Altman2015}. Also poorly understood is the problem of prethermalization in disorder, which did not receive much attention so far. In this article, we tackle this issue by studying the postquench prethermalization of a disordered quantum fluid of light.

Our system is based on the quantum propagation of a paraxial beam of quasimonochromatic light in a dispersive, inhomogeneous, and nonlinear dielectric medium. In this all-optical platform, the space propagation of the envelope of the quantum electric field may be reformulated in terms of the time evolution of a quantum fluid of interacting photons with specific canonical commutation relations \cite{Larre2015} (see also Refs.~\cite{Lai1989-a, Lai1989-b, Wright1991, Crosignani1995, Hagelstein1996, Kolobov1999, Matsko2000, Tsang2006, Drummond2004} for related works, especially in fiber geometries). The resulting analog system constitutes a particular class of ``quantum fluids of light'' \cite{Carusotto2013} and presently attracts a growing interest as a powerful tool for quantum simulating systems of many interacting particles \cite{Noh2017}. In close relation to the topic of the present paper, it was used to investigate the prethermalization \cite{Larre2016}, the thermalization, and the Bose-Einstein condensation \cite{Chiocchetta2016-a} of a homogeneous beam of quantum light, as well as their classical counterparts in a nonquantum description of the optical field \cite{Connaughton2005, Picozzi2007, Lagrange2007, Picozzi2008-a, Picozzi2008-b, Barviau2008, Barviau2009, Suret2010, Aschieri2011, Michel2011, Sun2012, Picozzi2012, Santic2018}.

\begin{figure}[t!]
\includegraphics[width=\linewidth]{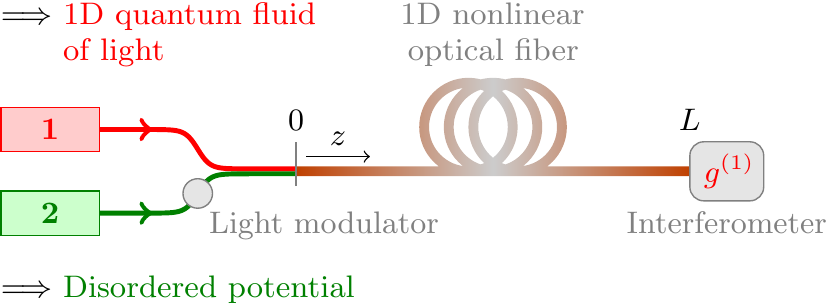}
\caption{(Color online) Schematics of the all-optical setup considered in the article (see the text).}
\label{Fig:Setup}
\end{figure}

The system specifically studied in this article is sketched in Fig.~\ref{Fig:Setup}. Two beams of light \maths{\boldsymbol{1}} and \maths{\boldsymbol{2}} copropagate in the positive-\maths{z} direction along a one-dimensional (1D) nonlinear optical fiber. In the fiber (\maths{0<z<L}), the two beams interact via the optical nonlinearity through cross-phase modulation, as detailed in Sec.~\ref{Sec:CrossPhaseModulation}. In this configuration, we demonstrate in Sec.~\ref{Sec:DisorderedOneDimensionalQuantumFluidOfLight} that the propagation of \maths{\boldsymbol{1}} in the presence of \maths{\boldsymbol{2}} may be reformulated in terms of the evolution of a disordered 1D quantum fluid of light. The beam \maths{\boldsymbol{1}} is treated within a quantum framework where the propagation coordinate \maths{z} and the time parameter \maths{t} play exchanged roles and where chromatic dispersion and self-phase modulation respectively provide an effective mass and effective two-body interactions to the photons. The power of the beam \maths{\boldsymbol{2}}, which enters the dynamics of the beam \maths{\boldsymbol{1}} through cross-phase modulation, acts as a disordered potential once randomized as a function of the spacelike variable \maths{t} by means of a light modulator. A Bogoliubov-type description of the vacuum fluctuations of the disordered 1D quantum fluid of light is provided in Sec.~\ref{Sec:QuantumDensityPhaseBogoliubovTheory}. When entering the fiber from free space (\maths{z=0}), the optical nonlinearity is abruptly switched on. As explained in Sec.~\ref{Sec:QuantumQuench}, this effectively simulates both a disorder and an interaction quench for the quantum fluid of light. The statistical properties of the postquench quantum fluid of light are encoded in its coherence function \maths{g^{(1)}}, which is the core object we study in the present work. In practice, \maths{g^{(1)}} can be experimentally accessed through interferometric measurements at the exit of the fiber (\maths{z=L}), and accordingly, we calculate it at this point in Sec.~\ref{Sec:PostquenchCoherence}. From its structure, we find that, as a result of the quench, a disorder-altered prethermalized state emerges in a light-cone way in the system, accompanied with a loss of macroscopic coherence. In Sec.~\ref{Sec:OnTheQuantumNatureOfTheDecoherenceAndOrdersOfMagnitude}, we argue on the quantum nature of our results and provide orders of magnitude based on state-of-the-art fiber optics. We conclude in Sec.~\ref{Sec:Conclusion}, after which technical points are collected in Appendices \ref{App:CrossPhaseModulationFromASingleBeamEffectiveModel} and \ref{App:FluxOfThePoyntingVectorOfACrossPhaseModulatedOpticalBeam}.

\section{Cross-phase modulation}
\label{Sec:CrossPhaseModulation}

In this section, we introduce the phenomenon of cross-phase modulation used in Sec.~\ref{Sec:DisorderedOneDimensionalQuantumFluidOfLight} to create an effective disordered potential for photons, and give the main hypotheses of our approach.

Two beams of light \maths{\boldsymbol{1}} and \maths{\boldsymbol{2}} copropagate in the positive-\maths{z} direction along a 1D optical fiber. Polarization effects are neglected so that a scalar description may be used. We express \maths{\boldsymbol{\alpha}}'s complex electric field \maths{E_{\alpha}(\mathbf{r},t)} [\maths{\alpha\in\{1,2\}} and \maths{\mathbf{r}=(x,y,z)}] as the product of an envelope \maths{\mathcal{E}_{\alpha}(\mathbf{r},t)} and a carrier \maths{e^{i(k_{\alpha}z-\omega_{\alpha}t)}} with propagation constant \maths{k_{\alpha}>0} and angular frequency \maths{\omega_{\alpha}}; we also split \maths{\mathcal{E}_{\alpha}(\mathbf{r},t)} into a transverse modal function \maths{F_{\alpha}(x,y)}, such that \maths{\int dx\,dy\,|F_{\alpha}(x,y)|^{2}=1}, times a longitudinal amplitude \maths{A_{\alpha}(z,t)}:
\begin{subequations}
\label{Eq:ElectricField}
\begin{align}
\label{Eq:ElectricField-a}
E_{\alpha}(\mathbf{r},t)&=\mathcal{E}_{\alpha}(\mathbf{r},t)\,e^{i(k_{\alpha}z-\omega_{\alpha}t)}, \\
\label{Eq:ElectricField-b}
\mathcal{E}_{\alpha}(\mathbf{r},t)&=F_{\alpha}(x,y)\,A_{\alpha}(z,t).
\end{align}
\end{subequations}
We now suppose quasimonochromaticity around \maths{\omega_{\alpha}}. In this case, \maths{A_{\alpha}(z,t)} becomes a slowly varying function of \maths{z} and \maths{t} over scales respectively of the order of \maths{2\pi/k_{\alpha}} and \maths{2\pi/\omega_{\alpha}}. Around \maths{\omega_{\alpha}}, the fiber displays a local Kerr nonlinearity of coefficient \maths{n_{2}(\omega)\lessgtr0} as well as the quadratic dispersion relation
\begin{equation}
\label{Eq:DispersionRelation}
k(\omega)\simeq k_{\alpha}+\frac{1}{v_{\alpha}}\,(\omega-\omega_{\alpha})+\frac{D_{\alpha}}{2}\,(\omega-\omega_{\alpha})^{2}.
\end{equation}
In this equation, \maths{v_{\alpha}=[(\partial k/\partial\omega)(\omega_{\alpha})]^{-1}>0} and \maths{D_{\alpha}=(\partial^{2}k/\partial\omega^{2})(\omega_{\alpha})\lessgtr0} are respectively the group velocity and the group-velocity-dispersion parameter at \maths{\omega_{\alpha}}. We finally assume negligible propagation losses at \maths{\omega_{\alpha}}.

In such a configuration, \maths{A_{\alpha}(z,t)} satisfies the following nonlinear propagation equation \cite{Agrawal2013}:
\begin{align}
\notag
i\,\frac{\partial A_{\alpha}}{\partial z}&\left.=\frac{D_{\alpha}}{2}\,\frac{\partial^{2}A_{\alpha}}{\partial t^{2}}-\frac{i}{v_{\alpha}}\,\frac{\partial A_{\alpha}}{\partial t}-\gamma_{\alpha}\,|A_{\alpha}|^{2}\,A_{\alpha}\right. \\
\label{Eq:PropagationEquation}
&\left.\hphantom{=}-\delta_{\alpha}\,|A_{3-\alpha}(z,t)|^{2}\,A_{\alpha}.\right.
\end{align}
In this equation, the nonlinear parameters \maths{\gamma_{\alpha}} and \maths{\delta_{\alpha}} are expressed as \maths{\gamma_{\alpha}=(\omega_{\alpha}/c_{0})\,(n_{2})_{\alpha}/\mathcal{A}_{\alpha}} and \maths{\delta_{\alpha}=2\,(\omega_{\alpha}/c_{0})\,(n_{2})_{\alpha}/\mathcal{A}}, with \maths{c_{0}} the speed of light in free space, \maths{(n_{2})_{\alpha}=n_{2}(\omega_{\alpha})} the Kerr-nonlinearity coefficient at \maths{\omega_{\alpha}}, \maths{\mathcal{A}_{\alpha}=[\int dx\,dy\,|F_{\alpha}(x,y)|^{4}]^{-1}} the effective transverse area of the beam of light \maths{\boldsymbol{\alpha}}, and \maths{\mathcal{A}=[\int dx\,dy\,|F_{1}(x,y)|^{2}\,|F_{2}(x,y)|^{2}]^{-1}} the overlap effective transverse area between \maths{\boldsymbol{1}} and \maths{\boldsymbol{2}}. Both the \maths{\gamma_{\alpha}} and the \maths{\delta_{\alpha}} terms in Eq.~\eqref{Eq:PropagationEquation} originate from the Kerr nonlinearity: While the \maths{\gamma_{\alpha}} term standardly describes self-phase modulation, the \maths{\delta_{\alpha}} one is as for it responsible for cross-phase modulation, nonlinear effect where one of the optical beams is phase-affected by its copropagating partner via the Kerr nonlinearity \cite{Agrawal2013, Chraplyvy1984, Alfano1986, Islam1987, Alfano1987}. The factor \maths{2} in \maths{\delta_{\alpha}} shows that cross-phase modulation is twice as effective as self-phase modulation when \maths{\mathcal{A}_{\alpha}\sim\mathcal{A}} and \maths{P_{1}\sim P_{2}}, where \maths{P_{\alpha}(z,t)=\frac{1}{2}\,c_{0}\,\varepsilon_{0}\,(n_{\mathrm{L}})_{\alpha}\,|A_{\alpha}(z,t)|^{2}} is the local and instantaneous power of the optical beam \maths{\boldsymbol{\alpha}}, with \maths{\varepsilon_{0}} the permittivity of free space and \maths{(n_{\mathrm{L}})_{\alpha}=n_{\mathrm{L}}(\omega_{\alpha})} the homogeneous contribution to the linear refractive index at \maths{\omega_{\alpha}}.

Note that Eq.~\eqref{Eq:PropagationEquation} may be derived within the framework of a phenomenological model where the cross-phase modulation induced by the beam of light \maths{\boldsymbol{3-\alpha}=\boldsymbol{2}~\text{or}~\boldsymbol{1}} is described by a modified linear refractive index for the single beam of light \maths{\boldsymbol{\alpha}=\boldsymbol{1}~\text{or}~\boldsymbol{2}}. This approach, implicitly used throughout this work, is detailed in Appendix \ref{App:CrossPhaseModulationFromASingleBeamEffectiveModel}.

\section{Disordered 1D quantum fluid of light}
\label{Sec:DisorderedOneDimensionalQuantumFluidOfLight}

In view of the discussions of Secs.~\ref{SubSec:QuantumNonlinearSchrodingerTheory} and \ref{SubSec:DisorderedPotential}, the coupled system formed by Eq.~\eqref{Eq:PropagationEquation} for \maths{\alpha=1} and by this same equation for \maths{\alpha=2} can be rearranged as
\begin{subequations}
\label{Eq:PropagationEquations}
\begin{align}
\notag
i\,\frac{\partial A_{1}}{\partial z}+\frac{i}{v_{1}}\,\frac{\partial A_{1}}{\partial t}&\left.=\frac{D_{1}}{2}\,\frac{\partial^{2}A_{1}}{\partial t^{2}}-\delta_{1}\,|A_{2}(z,t)|^{2}\,A_{1}\right. \\
\label{Eq:PropagationEquations-a}
&\left.\hphantom{=}-\gamma_{1}\,|A_{1}|^{2}\,A_{1},\right. \\
\notag
i\,\frac{\partial A_{2}}{\partial z}+\frac{i}{v_{2}}\,\frac{\partial A_{2}}{\partial t}&\left.=\frac{D_{2}}{2}\,\frac{\partial^{2} A_{2}}{\partial t^{2}}-\gamma_{2}\,|A_{2}|^{2}\,A_{2}\right. \\
\label{Eq:PropagationEquations-b}
&\left.\hphantom{=}-\delta_{2}\,|A_{1}(z,t)|^{2}\,A_{2}.\right.
\end{align}
\end{subequations}
In Sec.~\ref{SubSec:QuantumNonlinearSchrodingerTheory} first, we will present a formalism making it possible to describe the propagation \eqref{Eq:PropagationEquations-a} of the beam \maths{\boldsymbol{1}} at the quantum level. In Sec.~\ref{SubSec:DisorderedPotential} then, we will consider that the instantaneous power of the beam \maths{\boldsymbol{2}} is randomized by means of a light modulator (see Fig.~\ref{Fig:Setup}) so to produce disorder for the beam \maths{\boldsymbol{1}}. For this purpose, \maths{\boldsymbol{2}} will be treated within a nonquantum framework and in a configuration where its propagation \eqref{Eq:PropagationEquations-b} formally decouples from \eqref{Eq:PropagationEquations-a}.

\subsection{1D quantum nonlinear Schr\"odinger theory}
\label{SubSec:QuantumNonlinearSchrodingerTheory}

Capturing features originating from the zero-point fluctuations of the electric field of the optical beam \maths{\boldsymbol{1}} requires to build upon a quantum-field description of its propagation along the optical fiber. Following Refs.~\cite{Lai1989-a, Lai1989-b, Wright1991, Crosignani1995, Hagelstein1996, Kolobov1999, Matsko2000, Tsang2006}, a generalized quantum formulation of the propagation of a paraxial beam of quasimonochromatic scalar light in a dispersive, inhomogeneous, and nonlinear medium was derived from microscopic grounds in Ref.~\cite{Larre2015} and dimensionally reduced soon after, in Ref.~\cite{Larre2016}, to the nonlinear-optical-fiber geometry that interests us here. We assume that the envelope \maths{\mathcal{E}_{1}(\mathbf{r},t)} of \maths{\boldsymbol{1}}'s complex electric field propagates in the positive-\maths{z} direction (no back-propagating waves) but let it be arbitrarily (red- or blue-) detuned from the carrier angular frequency \maths{\omega_{1}}. In other words, the variables conjugated to \maths{\mathcal{E}_{1}(\mathbf{r},t)}'s variables \maths{z} and \maths{t} respectively take their values in \maths{(0,\infty)} and \maths{(-\infty,\infty)}. The first one, proportional to the linear momentum (along the \maths{z} axis) carried by \maths{\mathcal{E}_{1}(\mathbf{r},t)}, is bounded from below while the second one, proportional to the energy carried by \maths{\mathcal{E}_{1}(\mathbf{r},t)}, is not. As a result, \maths{\mathcal{E}_{1}(\mathbf{r},t)}'s variables \maths{z} and \maths{t} respectively behave as a time parameter and a space coordinate in the standard framework of quantum mechanics. Accordingly, within the single-beam effective model introduced in the end of Sec.~\ref{Sec:CrossPhaseModulation}, the canonical quantization procedure developed in Refs.~\cite{Larre2015, Larre2016} applies to the classical field \maths{\mathcal{E}_{1}(\mathbf{r},t)} \cite{Larre2015} and then to its longitudinal component \maths{A_{1}^{\vphantom{\ast}}(z,t)=\int dx\,dy\,F_{1}^{\ast}(x,y)\,\mathcal{E}_{1}^{\vphantom{\ast}}(\mathbf{r},t)} \cite{Larre2016}. Precisely, the latter is canonically replaced with a quantum field \maths{\hat{A}_{1}(z,t)} satisfying
\begin{subequations}
\label{Eq:QuantumFormalism}
\begin{align}
\notag
i\,\frac{\partial\hat{A}_{1}}{\partial z}+\frac{i}{v_{1}}\,\frac{\partial\hat{A}_{1}}{\partial t}&\left.=\frac{D_{1}}{2}\,\frac{\partial^{2}\hat{A}_{1}}{\partial t^{2}}-\delta_{1}\,|A_{2}(z,t)|^{2}\,\hat{A}_{1}\right. \\
\label{Eq:QuantumFormalism-a}
&\left.\hphantom{=}-\gamma_{1}^{\vphantom{\dag}}\,\hat{A}_{1}^{\dag}\,\hat{A}_{1}^{\vphantom{\dag}}\,\hat{A}_{1}^{\vphantom{\dag}},\right. \\
\label{Eq:QuantumFormalism-b}
[\hat{A}_{1}^{\vphantom{\dag}}(z,t),\hat{A}_{1}^{\dag}(z,t')]&\left.=\frac{\hbar}{\mathscr{C}}\,\delta(t-t'),\right. \\
\label{Eq:QuantumFormalism-c}
\mathscr{C}&\left.=\frac{1}{2}\,\frac{c_{0}\,\varepsilon_{0}\,(n_{\mathrm{L}})_{1}}{\omega_{1}}.\right.
\end{align}
\end{subequations}
The propagation equation \eqref{Eq:QuantumFormalism-a} is nothing but the quantized version of Eq.~\eqref{Eq:PropagationEquations-a} and the capacitance \eqref{Eq:QuantumFormalism-c} appearing in the same-``time \maths{z},'' different-``position \maths{t}'' commutation relation \eqref{Eq:QuantumFormalism-b} fixes the actual spacing between the accessible energy levels of the system \cite{Larre2015, Larre2016}.

The quantum theory \eqref{Eq:QuantumFormalism} is formally analogous to the one of dilute atomic Bose gases \cite{Dalfovo1999, Pitaevskii2016} in one dimension after exchanging the roles played by the position coordinate \maths{z} and the time parameter \maths{t}. Most particularly, apart from the constant-drift term \maths{(i/v_{1})\,\partial\hat{A}_{1}/\partial t} at the group velocity \maths{v_{1}}, Eq.~\eqref{Eq:QuantumFormalism-a} looks closely like the quantum nonlinear Schr\"odinger equation describing the dynamics of these atomic systems: \maths{\hat{A}_{1}(z,t)} corresponds to the quantum matter field in one dimension; \maths{-1/D_{1}} is the analog of the atom mass; \maths{-\delta_{1}\,|A_{2}(z,t)|^{2}} plays the role of an external potential; \maths{-\gamma_{1}} finally corresponds to the 1D atom-atom interaction constant in the zero-range-pseudopotential approximation. Noticeably, the incident quantum light field \maths{\hat{A}_{1}(0,t)} determines the initial condition of the quantum nonlinear Schr\"odinger-type equation \eqref{Eq:QuantumFormalism-a}, of first order in the partial derivative with respect to the timelike parameter \maths{z}. These analogies make it possible to reformulate the quantum propagation of the beam \maths{\boldsymbol{1}} in the presence of the beam \maths{\boldsymbol{2}} in terms of the evolution of a 1D quantum fluid of light in any external potential. This is what we detail in the next paragraphs.

To facilitate \maths{z} and \maths{t} to be viewed as time and space variables, we introduce
\begin{align}
\label{Eq:TimeVariable}
\tau&=\frac{z}{v_{1}}, \\
\label{Eq:SpaceVariable}
\zeta&=v_{1}\,t-z,
\end{align}
respectively homogeneous to a time and a length. In this new coordinate system, after defining
\begin{align}
\label{Eq:MatterField}
\hat{\Psi}(\zeta,\tau)&=\bigg(\frac{\mathscr{C}}{\hbar\,v_{1}}\bigg)^{\frac{1}{2}}\,\hat{A}_{1}\bigg(v_{1}\,\tau,\frac{\zeta}{v_{1}}+\tau\bigg), \\
\label{Eq:Mass}
m&=-\frac{\hbar}{v_{1}^{3}\,D_{1}^{\vphantom{3}}}, \\
\label{Eq:ExternalPotential}
U(\zeta,\tau)&=-\hbar\,v_{1}\,\delta_{1}\,\bigg|A_{2}\bigg(v_{1}\,\tau,\frac{\zeta}{v_{1}}+\tau\bigg)\bigg|^{2}, \\
\label{Eq:InteractionConstant}
g&=-\frac{(\hbar\,v_{1})^{2}\,\gamma_{1}}{\mathscr{C}},
\end{align}
the formalism \eqref{Eq:QuantumFormalism} explicitly takes the form of a 1D quantum nonlinear Schr\"odinger theory:
\begin{subequations}
\label{Eq:QuantumNLSFormalism}
\begin{gather}
\label{Eq:QuantumNLSFormalism-a}
i\,\hbar\,\frac{\partial\hat{\Psi}}{\partial\tau}=-\frac{\hbar^{2}}{2\,m}\,\frac{\partial^{2}\hat{\Psi}}{\partial\zeta^{2}}+U(\zeta,\tau)\,\hat{\Psi}+g\,\hat{\Psi}^{\dag}\,\hat{\Psi}\,\hat{\Psi}, \\
\label{Eq:QuantumNLSFormalism-b}
{\qquad}[\hat{\Psi}(\zeta,\tau),\hat{\Psi}^{\dag}(\zeta',\tau)]=\delta(\zeta-\zeta').{\qquad}
\end{gather}
\end{subequations}
Using \maths{\hat{\Psi}^{\dag}(\zeta,\tau)\,\hat{\Psi}(\zeta,\tau)=\hat{\Psi}(\zeta,\tau)\,\hat{\Psi}^{\dag}(\zeta,\tau)-\delta(0)} [from Eq.~\eqref{Eq:QuantumNLSFormalism-b}] and performing the substitution
\begin{equation}
\label{Eq:FirstSubstitution}
\hat{\Psi}(\zeta,\tau)\longrightarrow\hat{\Psi}(\zeta,\tau)\exp\!\bigg[i\,\frac{g\,\delta(0)\,\tau}{\hbar}\bigg],
\end{equation}
we rewrite the quantum nonlinear Schr\"odinger equation in the form \eqref{Eq:QuantumNLSFormalism-a} but with the interaction term replaced with \maths{g\,\hat{\Psi}\,\hat{\Psi}^{\dag}\,\hat{\Psi}} while preserving the same-\maths{\tau} commutation relation \eqref{Eq:QuantumNLSFormalism-b}. In doing so, the phase of \maths{\hat{\Psi}(\zeta,\tau)} conveniently disappears from the interaction term after reformulating the problem within Madelung's approach of quantum mechanics (see Sec.~\ref{Sec:QuantumDensityPhaseBogoliubovTheory}). Building upon \eqref{Eq:QuantumNLSFormalism} and \eqref{Eq:FirstSubstitution}, we will from now on make use of the terminology as well as of the theoretical tools specific to the physics of dilute atomic Bose gases.

At this stage, a few comments are in order. Note first that the extra \maths{z} dependence of \maths{\zeta} in Eq.~\eqref{Eq:SpaceVariable} makes the drift derivative \maths{(i/v_{1})\,\partial/\partial t=i\,\partial/\partial\zeta} disappear from the left-hand side of Eq.~\eqref{Eq:QuantumFormalism-a}. This is natural since Eq.~\eqref{Eq:SpaceVariable} links the coordinate systems \maths{\{x,y,z,t\}} and \maths{\{x,y,-\zeta,t\}} of two Galilean reference frames, the latter uniformly moving with respect to the former at the velocity \maths{v_{1}>0} along the \maths{z} axis. Second, the quantum field \eqref{Eq:MatterField}, \eqref{Eq:FirstSubstitution}, which describes the dynamics of the thus-defined 1D quantum fluid of light, is normalized so that the squared modulus of its classical version, \maths{|\Psi|^{2}}, coincides with the local and instantaneous density \maths{\rho} of photons in the beam \maths{\boldsymbol{1}}. Indeed, the flux \maths{\phi_{1}} of photons in that beam is by definition related to the density \maths{\rho} through \maths{\phi_{1}=v_{1}\,\rho} and to the power \maths{P_{1}=\frac{1}{2}\,c_{0}\,\varepsilon_{0}\,(n_{\mathrm{L}})_{1}\,|A_{1}|^{2}} through \maths{\phi_{1}=P_{1}/(\hbar\,\omega_{1})}, from which we get \maths{\rho=|\Psi|^{2}}, by definition of \maths{\Psi}. As a result, its quantized counterpart \maths{\hat{\rho}(\zeta,\tau)=\hat{\Psi}^{\dag}(\zeta,\tau)\,\hat{\Psi}(\zeta,\tau)} exactly corresponds to the density operator of the 1D quantum fluid of light. The mass \eqref{Eq:Mass} stems as for it from the chromatic dispersion of the optical fiber and may be positive or negative depending on whether the group-velocity dispersion is anomalous or normal at \maths{\omega_{1}}: \maths{m\gtrless0} when \maths{D_{1}\lessgtr0}. Finally, the external potential \eqref{Eq:ExternalPotential} and the photon-photon interactions, controlled by the nonlinear parameter \eqref{Eq:InteractionConstant}, originate from the Kerr nonlinearity and may be repulsive or attractive depending on whether the latter is defocusing or focusing at \maths{\omega_{1}}: \maths{U(\zeta,\tau),g\gtrless0} when \maths{(n_{2})_{1}\lessgtr0}. In fact, the 1D quantum fluid of light is robust against the formation of modulational instabilities when \maths{m} and \maths{g} are of same sign \cite{Larre2017}, for instance when they are both positive, \maths{m>0} (\maths{D_{1}<0}) and \maths{g>0} [\maths{(n_{2})_{1}<0}], which we consider from now on.

\subsection{Disordered potential}
\label{SubSec:DisorderedPotential}

As we eventually wish to describe the effect of a static disordered potential on the quantum fluid of light, we consider the particular case where \maths{U(\zeta,\tau)} only depends on the space coordinate \maths{\zeta}:
\begin{equation}
\label{Eq:StationaryExternalPotential}
U(\zeta,\tau)=V(\zeta).
\end{equation}
By looking at the propagation equation \eqref{Eq:PropagationEquations-b} of the optical field \maths{A_{2}(z,t)}---from the squared modulus of which \maths{U(\zeta,\tau)} is determined [see Eq.~\eqref{Eq:ExternalPotential}]---, the condition \eqref{Eq:StationaryExternalPotential} may be achieved in a configuration where (i) \maths{v_{2}=v_{1}}, (ii) \maths{D_{2}=0}, and (iii) \maths{(n_{2})_{2}=0}, the latter constraint yielding \maths{\gamma_{2}=0} and \maths{\delta_{2}=0}. Indeed, in this very particular case, \maths{A_{2}(z,t)} is reduced to obey the simple propagation equation
\begin{equation}
\label{Eq:TwoCoreFiber}
\frac{\partial A_{2}}{\partial z}+\frac{1}{v_{1}}\,\frac{\partial A_{2}}{\partial t}=0,
\end{equation}
the solutions of which are by construction functions of \maths{v_{1}\,t-z=\zeta}. From an experimental point of view, the conditions (i)--(iii) could be realized in the following configuration. The optical fiber is designed so to have two distinct cores. The first core supports the beam \maths{\boldsymbol{1}} while the beam \maths{\boldsymbol{2}} propagates along the second core. The latter is made of a linear material [condition (iii)] whose dispersion relation \maths{\kappa(\omega)} is such that the group velocity \maths{[(\partial\kappa/\partial\omega)(\omega)]^{-1}} at \maths{\omega=\omega_{2}} equals the group velocity \maths{v_{1}} in the first core [condition (i)] and such that the group-velocity-dispersion parameter \maths{(\partial^{2}\kappa/\partial\omega^{2})(\omega)} vanishes at \maths{\omega=\omega_{2}}, i.e., such that \maths{\lambda_{\mathrm{D}}=2\pi\,c_{0}/\omega_{2}} corresponds to the so-called zero-dispersion wavelength \cite{Agrawal2013} of the material [condition (ii)]. By construction physically separated from \maths{\boldsymbol{1}}, the beam \maths{\boldsymbol{2}} should nevertheless be sufficiently evanescent in the \maths{x} and \maths{y} directions to make \maths{\boldsymbol{1}} interact with it through cross-phase modulation [that is, to always have Eq.~\eqref{Eq:PropagationEquations-a}]. Noticeably, since \maths{\boldsymbol{2}} is here assumed to propagate in a linear material, cross-phase modulation does not enter the dynamics of \maths{A_{2}(z,t)}, the propagation equation of which is then decoupled from Eq.~\eqref{Eq:PropagationEquations-a}.

In this work, we are interested in describing the vacuum fluctuations of the 1D quantum fluid of light in the presence of disorder (Sec.~\ref{Sec:QuantumDensityPhaseBogoliubovTheory}). Such a disorder can be obtained by tailoring the stationary external potential \eqref{Eq:StationaryExternalPotential} so that it becomes a random function of \maths{\zeta}. Since \maths{V(\zeta)} is derived from \maths{|A_{2}(z,t)|^{2}} and since \maths{\zeta} is nothing but \maths{t} at a fixed \maths{z}, this may be achieved by randomly designing the input power profile \maths{P_{2}(0,t)=\frac{1}{2}\,c_{0}\,\varepsilon_{0}\,(n_{\mathrm{L}})_{2}\,|A_{2}(0,t)|^{2}} of the optical beam \maths{\boldsymbol{2}} as a function of \maths{t}, typically making use of a light modulator (see Fig.~\ref{Fig:Setup}). From now on, we denote by \maths{\overline{\cdots}} the average over the realizations of the disorder. For the sake of convenience, we perform the gauge transformations
\begin{align}
\label{Eq:SecondSubstitution}
V(\zeta)&\longrightarrow V(\zeta)-\overline{V(\zeta)}, \\
\label{Eq:ThirdSubstitution}
\hat{\Psi}(\zeta,\tau)&\longrightarrow\hat{\Psi}(\zeta,\tau)\exp\!\bigg[i\,\frac{\overline{V(\zeta)}\,\tau}{\hbar}\bigg]
\end{align}
in Eqs.~\eqref{Eq:QuantumNLSFormalism} supplemented by \eqref{Eq:FirstSubstitution} and \eqref{Eq:StationaryExternalPotential}. As a consequence, we are led to investigate the following 1D quantum nonlinear Schr\"odinger problem:
\begin{subequations}
\label{Eq:QuantumNLSFormalismBis}
\begin{gather}
\label{Eq:QuantumNLSFormalismBis-a}
i\,\hbar\,\frac{\partial\hat{\Psi}}{\partial\tau}=-\frac{\hbar^{2}}{2\,m}\,\frac{\partial^{2}\hat{\Psi}}{\partial\zeta^{2}}+V(\zeta)\,\hat{\Psi}+g\,\hat{\Psi}\,\hat{\Psi}^{\dag}\,\hat{\Psi}, \\
\label{Eq:QuantumNLSFormalismBis-b}
{\qquad}[\hat{\Psi}(\zeta,\tau),\hat{\Psi}^{\dag}(\zeta',\tau)]=\delta(\zeta-\zeta'),{\qquad}
\end{gather}
\end{subequations}
where the static disordered potential \maths{V(\zeta)} is now of zero average:
\begin{equation}
\label{Eq:DisorderAverage}
\overline{V(\zeta)}=0.
\end{equation}
In the following, we will also need its two-point correlation function, which we choose to be Gaussian:
\begin{equation}
\label{Eq:DisorderCorrelations}
\overline{V(\zeta)\,V(\zeta')}=\mathcal{V}^{2}\,C(\zeta-\zeta')=\mathcal{V}^{2}\,e^{-(\zeta-\zeta')^{2}/\sigma^{2}},
\end{equation}
where \maths{\mathcal{V}=[\overline{V^{2}(\zeta)}]^{1/2}} and \maths{\sigma} are respectively the standard deviation and the correlation length of \maths{V(\zeta)}. Note that since we will always work at the second order in \maths{\mathcal{V}} throughout this paper, our calculations will hold whatever the probability distribution of \maths{V(\zeta)}.

\section{Quantum Bogoliubov theory for disordered 1D systems}
\label{Sec:QuantumDensityPhaseBogoliubovTheory}

In order to perform an analytical treatment of the quantum dynamics \eqref{Eq:QuantumNLSFormalismBis} in the external potential \maths{V(\zeta)}, we assume that our nonlinear optical system falls into the limits of weak interactions and of small density quantum fluctuations. These hypotheses delimit the framework of the density-phase extension \cite{Petrov2003, Mora2003} of the well-known Bogoliubov theory of linearized quantum fluctuations \cite{Dalfovo1999, Pitaevskii2016, Bogoliubov1947}. It is standardly used to treat the infrared divergences of the phase fluctuations in reduced dimensions. In this section, we recall the main lines of this approach for our disordered 1D quantum fluid of light, taking inspiration from Refs.~\cite{Gaul2011, Gaul2010, Lugan2011, Lugan2010}. For the moment, we focus on light propagation in the fiber, leaving the question of the interfaces for the next section.

We start by writing the quantum field \maths{\hat{\Psi}(\zeta,\tau)} in Madelung's representation \cite{Petrov2003, Mora2003}:
\begin{equation}
\label{Eq:MadelungRepresentation}
\hat{\Psi}(\zeta,\tau)=e^{i\hat{\varphi}(\zeta,\tau)}\sqrt{\hat{\rho}(\zeta,\tau)},
\end{equation}
where the density and the phase Hermitian operators \maths{\hat{\rho}(\zeta,\tau)} and \maths{\hat{\varphi}(\zeta,\tau)} obey the commutation rule
\begin{equation}
\label{Eq:CommutationRelationDensityPhase}
[\hat{\rho}(\zeta,\tau),\hat{\varphi}(\zeta',\tau)]=i\,\delta(\zeta-\zeta')
\end{equation}
at any time \maths{\tau} so as to preserve the canonical commutation relation \eqref{Eq:QuantumNLSFormalismBis-b}. Inserting Eq.~\eqref{Eq:MadelungRepresentation} into Eq.~\eqref{Eq:QuantumNLSFormalismBis-a} and separating the imaginary parts from the real ones in the resulting Heisenberg equation of motion, we obtain the well-known Madelung (or quantum Euler) equations \cite{Petrov2003, Mora2003}
\begin{subequations}
\label{Eq:MadelungEquations}
\begin{gather}
\label{Eq:MadelungEquations-a}
\frac{\partial\hat{\rho}}{\partial\tau}+\frac{\partial}{\partial\zeta}(\hat{v}\,\hat{\rho})=0, \\
\label{Eq:MadelungEquations-b}
m\,\frac{\partial\hat{v}}{\partial\tau}=-\frac{\partial}{\partial\zeta}\bigg[\frac{m\,\hat{v}^{2}}{2}-\frac{\hbar^{2}}{2\,m}\,\frac{1}{\sqrt{\hat{\rho}}}\,\frac{\partial^{2}\sqrt{\hat{\rho}}}{\partial\zeta^{2}}+V(\zeta)+g\,\hat{\rho}\bigg]
\end{gather}
\end{subequations}
for the density and the velocity operators \maths{\hat{\rho}(\zeta,\tau)} and \maths{\hat{v}(\zeta,\tau)=(\hbar/m)\,(\partial\hat{\varphi}/\partial\zeta)(\zeta,\tau)}.

In 1D, the hypothesis of weak interactions implies the one of small density quantum fluctuations \cite{NoteBogoliubov}. Since the external potential is in addition time independent, we accordingly look for weak-amplitude quantum fluctuations of the density operator around a stationary classical state of density \maths{\rho_{0}(\zeta)}, zero velocity for simplicity's sake, and overall energy (chemical potential) \maths{\mu}:
\begin{align}
\label{Eq:QuantumFluctuationsDensity}
\hat{\rho}(\zeta,\tau)&=\rho_{0}(\zeta)+\hat{\rho}_{1}(\zeta,\tau), \\
\label{Eq:QuantumFluctuationsPhase}
\hat{\varphi}(\zeta,\tau)&=\varphi_{0}(\tau)+\hat{\varphi}_{1}(\zeta,\tau)=-\frac{\mu\,\tau}{\hbar}+\hat{\varphi}_{1}(\zeta,\tau).
\end{align}
In Eq.~\eqref{Eq:QuantumFluctuationsDensity}, the classical density \maths{\rho_{0}(\zeta)} a priori depends on \maths{\zeta} in the presence of the inhomogeneous potential \maths{V(\zeta)}. Its quantum correction \maths{\hat{\rho}_{1}(\zeta,\tau)} is in comparison small. In Eqs.~\eqref{Eq:QuantumFluctuationsPhase}, the classical phase \maths{\varphi_{0}(\tau)=-\mu\,\tau/\hbar} does not depend on \maths{\zeta} in the absence of background velocity. Its quantum correction \maths{\hat{\varphi}_{1}(\zeta,\tau)} strongly fluctuates in the infrared. This is not the case for the associated velocity field \maths{(\hbar/m)\,(\partial\hat{\varphi}_{1}/\partial\zeta)(\zeta,\tau)}, which is as weakly fluctuating as \maths{\hat{\rho}_{1}(\zeta,\tau)} \cite{Petrov2003, Mora2003}.

\subsection{Gross-Pitaevskii classical field}
\label{SubSec:GrossPitaevskiiClassicalField}

At the classical level, that is, when \maths{\hat{\rho}(\zeta,\tau)=\rho_{0}(\zeta)} and \maths{\hat{\varphi}(\zeta,\tau)=\varphi_{0}(\tau)=-\mu\,\tau/\hbar}, Eq.~\eqref{Eq:MadelungEquations-a} is trivially verified and Eq.~\eqref{Eq:MadelungEquations-b} simplifies to the following stationary Gross-Pitaevskii equation for the classical density \maths{\rho_{0}(\zeta)} \cite{Petrov2003, Mora2003}:
\begin{equation}
\label{Eq:GPE}
\mu=-\frac{\hbar^{2}}{2\,m}\,\frac{1}{\sqrt{\rho_{0}}}\,\frac{\partial^{2}\sqrt{\rho_{0}}}{\partial\zeta^{2}}+V(\zeta)+g\,\rho_{0}.
\end{equation}

In this work, we assume that the fiber is continuously pumped by a monochromatic beam \maths{\boldsymbol{1}} (the initial condition will be precisely treated in Sec.~\ref{SubSec:QuantumOpticalFieldInFreeSpace}, when dealing with the interfaces with free space). Therefore, in the absence of disorder [\maths{V(\zeta)=0}], the in-fiber classical density \maths{\rho_{0}(\zeta)} is externally forced to be independent of \maths{\zeta}, given by the uniform solution \maths{\bar{\rho}_{0}} of the stationary Gross-Pitaevskii equation \eqref{Eq:GPE}:
\begin{equation}
\label{Eq:HomogeneousDensity}
\rho_{0}(\zeta)=\bar{\rho}_{0}=\mathrm{const},\quad\text{with}\quad\mu=g\,\bar{\rho}_{0}.
\end{equation}

In the presence of disorder [\maths{V(\zeta)\neq0}], \maths{\rho_{0}(\zeta)} cannot be independent of \maths{\zeta} anymore. From now on, we assume that the typical amplitude \maths{[\overline{V^{2}(\zeta)}]^{1/2}=\mathcal{V}} of the disordered potential is much smaller than the typical interaction energy \maths{g\,\bar{\rho}_{0}=\mu}: \maths{\mathcal{V}\ll\mu}. In this small-disorder limit, \maths{\rho_{0}(\zeta)} weakly deviates from its disorder-average value \maths{\overline{\rho_{0}(\zeta)}=\bar{\rho}_{0}} as
\begin{equation}
\label{Eq:ClassicalFluctuationsDensity}
\rho_{0}(\zeta)=\bar{\rho}_{0}+\delta\rho_{0}(\zeta),
\end{equation}
where \maths{|\delta\rho_{0}(\zeta)|/\bar{\rho}_{0}\sim\mathcal{V}/\mu\ll1}. Note that due to the stationarity of the amplitude of the input classical beam, no localization phenomenon is visible in the in-fiber average classical density, which is purely uniform. The situation could be different for explicitly time-dependent, pulsed beams. After linearizing Eq.~\eqref{Eq:GPE} according to Eq.~\eqref{Eq:ClassicalFluctuationsDensity}, we get the linear differential equation
\begin{equation}
\label{Eq:LinearizedGPE}
\bigg({-}\frac{\xi^{2}}{4}\,\frac{\partial^{2}}{\partial\zeta^{2}}+1\bigg)\,\frac{\delta\rho_{0}}{\bar{\rho}_{0}}=-\frac{V(\zeta)}{\mu},
\end{equation}
where \maths{\xi=\hbar/(m\,\mu)^{1/2}} is the healing length. Solving it in Fourier space, we obtain \cite{Pavloff2002, SanchezPalencia2006, Paul2007}
\begin{subequations}
\label{Eq:SolutionLinearizedGPE}
\begin{align}
\label{Eq:SolutionLinearizedGPE-a}
\delta\rho_{0}(\zeta)&=\int d\zeta'\,\chi(\zeta-\zeta')\,V(\zeta'), \\
\label{Eq:SolutionLinearizedGPE-b}
\chi(\zeta-\zeta')&=-\frac{\bar{\rho}_{0}}{\xi\,\mu}\,e^{-2|\zeta-\zeta'|/\xi}.
\end{align}
\end{subequations}
Equation \eqref{Eq:SolutionLinearizedGPE-a} gives the density linear response of the 1D system, and the expression \eqref{Eq:SolutionLinearizedGPE-b} of the corresponding linear-response function unsurprisingly indicates that the typical length scale over which the fluid's density is able to respond to a single realization of the disorder is the healing length \maths{\xi}.

\begin{figure}[t!]
\includegraphics[width=\linewidth]{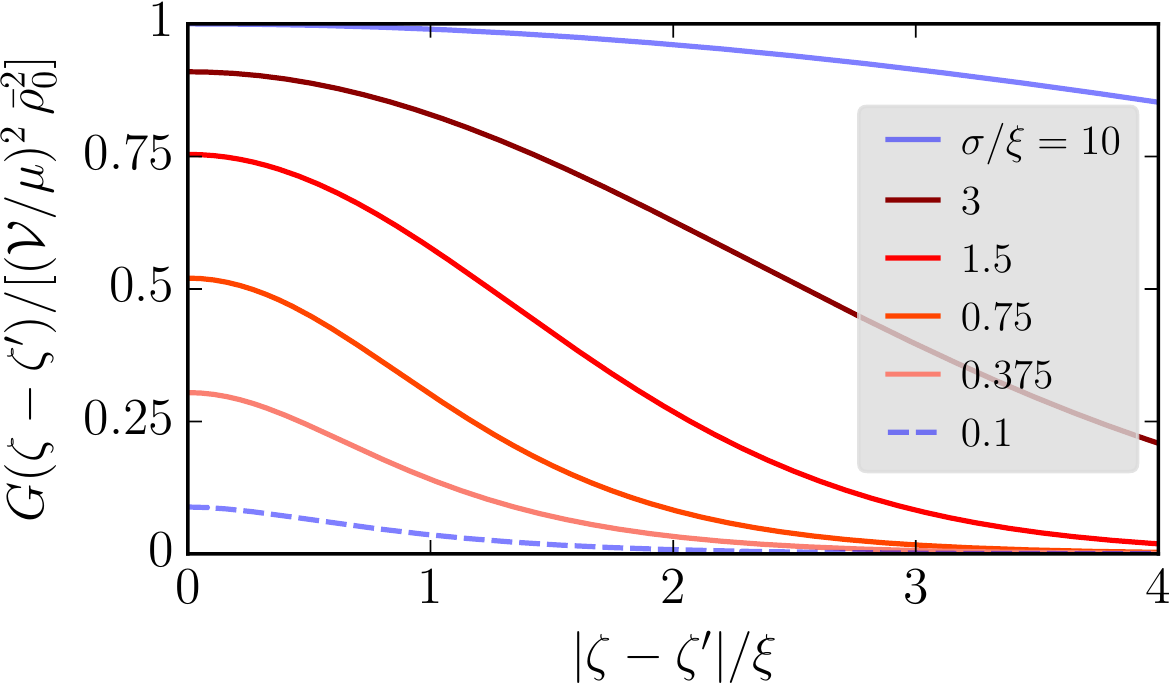}
\caption{(Color online) Red curves: Normalized two-point correlation function of the density classical fluctuations as a function of \maths{|\zeta-\zeta'|/\xi} for different values of \maths{\sigma/\xi}, as given in Eqs.~\eqref{Eq:DensityCorrelationsResult}. Blue curves: Asymptotic results when \maths{\sigma/\xi\ll1} (dashed curve) and when \maths{\sigma/\xi\gg1} (solid curve), as given in Eq.~\eqref{Eq:DensityCorrelationsAsymptoticResult}.}
\label{Fig:DensityCorrelations}
\end{figure}

Given Eq.~\eqref{Eq:ClassicalFluctuationsDensity} with \maths{|\delta\rho_{0}(\zeta)|/\bar{\rho}_{0}\sim\mathcal{V}/\mu\ll1}, any expectation value of quantities involving \maths{\rho_{0}(\zeta)} starts to depend on disorder from the second order when expanded in powers of \maths{\mathcal{V}/\mu}. We will work up to this order in the following, assuming that the subsequent terms, a priori smaller, do not alter the general physics of the problem. For this reason, it is sufficient to know the two-point correlator
\begin{equation}
\label{Eq:DensityCorrelations}
G(\zeta-\zeta')=\overline{\delta\rho_{0}(\zeta)\,\delta\rho_{0}(\zeta')},
\end{equation}
which only depends on \maths{|\zeta-\zeta'|} since \maths{\bar{\rho}_{0}=\mathrm{const}}. Making use of Eqs.~\eqref{Eq:DisorderCorrelations} and \eqref{Eq:SolutionLinearizedGPE}, we recast Eq.~\eqref{Eq:DensityCorrelations} as
\begin{align}
\notag
\frac{G(\zeta-\zeta')}{\bar{\rho}_{0}^{2}}&\left.=\bigg(\frac{\mathcal{V}}{\mu}\bigg)^{2}\int\frac{dZ\,dZ'}{\xi^{2}}\,C(Z-Z')\right. \\
\label{Eq:DensityCorrelationsBis}
&\left.\hphantom{=}\times e^{-2(|Z-\zeta|+|Z'-\zeta'|)/\xi},\right.
\end{align}
where \maths{C(Z-Z')=e^{-(Z-Z')^{2}/\sigma^{2}}}. Performing the integrals, we get
\begin{subequations}
\label{Eq:DensityCorrelationsResult}
\begin{align}
\notag
\frac{G(\zeta-\zeta')}{\bar{\rho}_{0}^{2}}&\left.=\frac{\sqrt{\pi}}{2}\,\bigg(\frac{\mathcal{V}}{\mu}\bigg)^{2}\,\frac{\sigma}{\xi}\right. \\
\notag
&\left.\hphantom{=}\times\bigg[e^{\sigma^{2}/\xi^{2}}\,\frac{f(\zeta-\zeta')+f(\zeta'-\zeta)}{2}\right. \\
\label{Eq:DensityCorrelationsResult-a}
&\left.\hphantom{=}+\frac{2}{\sqrt{\pi}}\,\frac{\sigma}{\xi}\,e^{-(\zeta-\zeta')^{2}/\sigma^{2}}\bigg],\right. \\
\notag
f(\zeta-\zeta')&\left.=\bigg(1-2\,\frac{\sigma^{2}}{\xi^{2}}+2\,\frac{\zeta-\zeta'}{\xi}\bigg)\,e^{-2(\zeta-\zeta')/\xi}\right. \\
\label{Eq:DensityCorrelationsResult-b}
&\left.\hphantom{=}\times\mathrm{erfc}\bigg(\frac{\sigma}{\xi}-\frac{\zeta-\zeta'}{\sigma}\bigg),\right.
\end{align}
\end{subequations}
where \maths{\mathrm{erfc}(X)=(2/\sqrt{\pi})\int_{X}^{\infty}dY\,e^{-Y^{2}}} is the complementary error function. In Fig.~\ref{Fig:DensityCorrelations}, we plot \maths{G(\zeta-\zeta')/[(\mathcal{V}/\mu)^{2}\,\bar{\rho}_{0}^{2}]} as a function of \maths{|\zeta-\zeta'|/\xi} for different values of \maths{\sigma/\xi}.

When \maths{\sigma/\xi\ll1} or \maths{\sigma/\xi\gg1}, Eqs.~\eqref{Eq:DensityCorrelationsResult} reduce to
\begin{equation}
\label{Eq:DensityCorrelationsAsymptoticResult}
\frac{G(\zeta-\zeta')}{\bar{\rho}_{0}^{2}}\simeq
\begin{dcases}
\frac{\sqrt{\pi}}{2}\,\bigg(\frac{\mathcal{V}}{\mu}\bigg)^{2}\,\frac{\sigma}{\xi} e^{-2|\zeta-\zeta'|/\xi}\\
\times\bigg(1+2\,\frac{|\zeta-\zeta'|}{\xi}\bigg) & \frac{\sigma}{\xi}\ll1 \\
\bigg(\frac{\mathcal{V}}{\mu}\bigg)^{2}\,e^{-(\zeta-\zeta')^{2}/\sigma^{2}} & \frac{\sigma}{\xi}\gg1.
\end{dcases}
\end{equation}
These asymptotic behaviors are shown in Fig.~\ref{Fig:DensityCorrelations}. In the limit \maths{\sigma/\xi\ll1}, the correlation function \maths{C(Z-Z')} in Eq.~\eqref{Eq:DensityCorrelationsBis} can be replaced by \maths{\sqrt{\pi}\,\sigma\,\delta(Z-Z')} (uncorrelated disorder). In this case, the healing length \maths{\xi} is the only relevant scale of variation of \maths{G(\zeta-\zeta')}. In the inverse limit \maths{\sigma/\xi\gg1}, \maths{C(Z-Z')} slowly varies at the scale of \maths{\xi}. In this case, \maths{G(\zeta-\zeta')} and \maths{C(\zeta-\zeta')} are proportional---interestingly independently of the shape of \maths{C(\zeta-\zeta')}---and the disorder's correlation length \maths{\sigma} is the only relevant scale of variation of \maths{G(\zeta-\zeta')}. The limit \maths{\sigma/\xi\gg1} actually coincides with the Thomas-Fermi regime where the kinetic term, \maths{\sim\hbar^{2}/(m\,\sigma^{2})}, is negligible compared to the interaction term, \maths{\sim\hbar^{2}/(m\,\xi^{2})}, in the Gross-Pitaevskii equation \eqref{Eq:GPE}. In this limit indeed, Eq.~\eqref{Eq:GPE} reduces to \maths{\rho_{0}(\zeta)\simeq[\mu-V(\zeta)]/g} (since \maths{\mu>\mathcal{V}}), from which we readily get the second row of Eq.~\eqref{Eq:DensityCorrelationsAsymptoticResult} after making use of Eq.~\eqref{Eq:ClassicalFluctuationsDensity}, \maths{\mu=g\,\bar{\rho}_{0}}, and Eqs.~\eqref{Eq:DisorderCorrelations}.

\subsection{Bogoliubov quantum fluctuations}
\label{SubSec:BogoliubovQuantumFluctuations}

At the first order in the density and the velocity quantum fluctuations, \maths{\hat{\rho}_{1}(\zeta,\tau)} and \maths{(\hbar/m)\,(\partial\hat{\varphi}_{1}/\partial\zeta)(\zeta,\tau)}, the Madelung equations \eqref{Eq:MadelungEquations} reduce to the density-phase Bogoliubov-de Gennes equations for the rescaled quantum fields \maths{\hat{\rho}_{1}(\zeta,\tau)/\sqrt{\rho_{0}(\zeta)}} and \maths{2\,i\,\sqrt{\rho_{0}(\zeta)}\;\hat{\varphi}_{1}(\zeta,\tau)} \cite{Petrov2003, Mora2003}:
\begin{subequations}
\label{Eq:BdGEquations}
\begin{align}
\notag
&\left.i\,\hbar\,\frac{\partial}{\partial\tau}\,\frac{\hat{\rho}_{1}}{\sqrt{\rho_{0}(\zeta)}}\right. \\
\label{Eq:BdGEquations-a}
&\left.{\quad}=\bigg[{-}\frac{\hbar^{2}}{2\,m}\,\frac{\partial^{2}}{\partial\zeta^{2}}+V(\zeta)+g\,\rho_{0}(\zeta)-\mu\bigg]\,2\,i\,\sqrt{\rho_{0}(\zeta)}\;\hat{\varphi}_{1},\right. \\
\notag
&\left.i\,\hbar\,\frac{\partial}{\partial\tau}\,2\,i\,\sqrt{\rho_{0}(\zeta)}\;\hat{\varphi}_{1}\right. \\
\label{Eq:BdGEquations-b}
&\left.{\quad}=\bigg[{-}\frac{\hbar^{2}}{2\,m}\,\frac{\partial^{2}}{\partial\zeta^{2}}+V(\zeta)+3\,g\,\rho_{0}(\zeta)-\mu\bigg]\,\frac{\hat{\rho}_{1}}{\sqrt{\rho_{0}(\zeta)}},\right.
\end{align}
\end{subequations}
where \maths{\rho_{0}(\zeta)} is given in Eqs.~\eqref{Eq:ClassicalFluctuationsDensity} and \eqref{Eq:SolutionLinearizedGPE} in the weak-disorder limit \maths{\mathcal{V}/\mu\ll1}.

Let us first recall a few well-known results in the absence of disorder [\maths{V(\zeta)=0}]. In this case, the system's background density is homogeneous, Eq.~\eqref{Eq:HomogeneousDensity}. As a result, the eigensolutions \maths{\hat{\rho}_{1}(\zeta,\tau)/\sqrt{\bar{\rho}_{0}}} and \maths{2\,i\,\sqrt{\bar{\rho}_{0}}\;\hat{\varphi}_{1}(\zeta,\tau)} of the Bogoliubov-de Gennes equations \eqref{Eq:BdGEquations} are linear superpositions of plane-wave fields:
\begin{equation}
\label{Eq:FourierTransformDensityPhase}
\begin{bmatrix}
\hat{\rho}_{1}(\zeta,\tau)/\sqrt{\bar{\rho}_{0}} \\
2\,i\,\sqrt{\bar{\rho}_{0}}\;\hat{\varphi}_{1}(\zeta,\tau)
\end{bmatrix}
=\int\frac{dk}{2\pi}
\begin{bmatrix}
\hat{\rho}_{1}(k,\tau)/\sqrt{\bar{\rho}_{0}} \\
2\,i\,\sqrt{\bar{\rho}_{0}}\;\hat{\varphi}_{1}(k,\tau)
\end{bmatrix}
e^{ik\zeta}.
\end{equation}
Their common wavenumber along the \maths{\zeta} axis is denoted by \maths{k} and is proportional to the detuning \maths{\Delta=\omega-\omega_{1}\lessgtr0} from the carrier angular frequency \maths{\omega_{1}} since \maths{\zeta} is nothing but the time variable \maths{t} (at a fixed \maths{z}) of the complex envelope \maths{\mathcal{E}_{1}(\mathbf{r},t)}. We furthermore parametrize their amplitudes as follows:
\begin{equation}
\label{Eq:BogoliubovRotation}
\begin{bmatrix}
\hat{\rho}_{1}(k,\tau)/\sqrt{\bar{\rho}_{0}} \\
2\,i\,\sqrt{\bar{\rho}_{0}}\;\hat{\varphi}_{1}(k,\tau)
\end{bmatrix}
=(u_{k}\pm v_{k})\,[\hat{a}(k,\tau)\pm\hat{a}^{\dag}(-k,\tau)].
\end{equation}
In doing this, \maths{\hat{\rho}_{1}(\zeta,\tau)/\sqrt{\bar{\rho}_{0}}} and \maths{2\,i\,\sqrt{\bar{\rho}_{0}}\;\hat{\varphi}_{1}(\zeta,\tau)} may be symmetrically expressed as
\begin{equation}
\label{Eq:HermitianAntiHermitianDecomposition}
\begin{bmatrix}
\hat{\rho}_{1}(\zeta,\tau)/\sqrt{\bar{\rho}_{0}} \\
2\,i\,\sqrt{\bar{\rho}_{0}}\;\hat{\varphi}_{1}(\zeta,\tau)
\end{bmatrix}
=\hat{\gamma}(\zeta,\tau)\pm\hat{\gamma}^{\dag}(\zeta,\tau),
\end{equation}
where
\begin{equation}
\label{Eq:BogoliubovFluctuation}
{\hat{\gamma}(\zeta,\tau)=\int\frac{dk}{2\pi}\,[u_{k}\,e^{ik\zeta}\,\hat{a}(k,\tau)+v_{k}\,e^{-ik\zeta}\,\hat{a}^{\dag}(k,\tau)].}
\end{equation}
In the transformation equation \eqref{Eq:BogoliubovRotation}, the operator \maths{\hat{a}(k,\tau)} [\maths{\hat{a}^{\dag}(-k,\tau)}] annihilates (creates) an elementary excitation in the plane-wave mode of wavenumber \maths{k} (\maths{-k}) at the time \maths{\tau}. It harmonically evolves at a well-defined energy \maths{E_{k}>0} and satisfies the standard equal-time commutation relation in momentum space:
\begin{align}
\label{Eq:PlaneWaveEvolution}
\hat{a}(k,\tau)&=e^{-iE_{k}\tau/\hbar}\,\hat{a}(k,0), \\
\label{Eq:CommutationRelationBogoliubovOperators}
[\hat{a}(k,\tau),\hat{a}^{\dag}(k',\tau)]&=2\pi\,\delta(k-k').
\end{align}
Chosen to be real and even functions of \maths{k}, the weights \maths{u_{k}} and \maths{v_{k}} must accordingly obey the constraint
\begin{equation}
\label{Eq:BogoliubovNormalization}
u_{k}^{2}-v_{k}^{2}=1
\end{equation}
so as to preserve the commutation rule \eqref{Eq:CommutationRelationDensityPhase} verified by the fields \maths{\hat{\rho}(\zeta,\tau)} and \maths{\hat{\varphi}(\zeta,\tau)}. When \maths{V(\zeta)=0}, Eqs.~\eqref{Eq:BdGEquations}, \eqref{Eq:BogoliubovRotation}, \eqref{Eq:PlaneWaveEvolution}, and \eqref{Eq:BogoliubovNormalization} are readily solved in Fourier space and give the standard results \cite{Dalfovo1999, Pitaevskii2016, Petrov2003, Mora2003, Bogoliubov1947}
\begin{align}
\label{Eq:BogoliubovSpectrum}
E_{k}&=\bigg[\frac{\hbar^{2}\,k^{2}}{2\,m}\,\bigg(\frac{\hbar^{2}\,k^{2}}{2\,m}+2\,\mu\bigg)\bigg]^{\frac{1}{2}}, \\
\label{Eq:BogoliubovAmplitudes}
u_{k}\pm v_{k}&=\bigg(\frac{\hbar^{2}\,k^{2}}{2\,m}\bigg/E_{k}\bigg)^{\pm\frac{1}{2}}.
\end{align}
Equation \eqref{Eq:BogoliubovSpectrum} is the usual Bogoliubov dispersion relation and Eq.~\eqref{Eq:BogoliubovAmplitudes} fixes the \maths{k} dependence of the Bogoliubov amplitudes \maths{u_{k}} and \maths{v_{k}}. When \maths{|k|\,\xi\ll1}, the Bogoliubov excitations of the homogeneous quantum fluid of light consist in sound waves, or phonons, propagating at the velocity \maths{s=(\mu/m)^{1/2}=\hbar/(m\,\xi)}: \maths{E_{k}\simeq s\,\hbar\,|k|}. When \maths{|k|\,\xi\gg1} instead, they consist in gapped free particles: \maths{E_{k}\simeq\hbar^{2}\,k^{2}/(2\,m)+\mu}.

Let us now come back to the situation where disorder is present [\maths{V(\zeta)\neq0}]. In this case, the system's background density \maths{\rho_{0}(\zeta)} is no longer constant. It is given in Eqs.~\eqref{Eq:ClassicalFluctuationsDensity} and \eqref{Eq:SolutionLinearizedGPE} in the weak-disorder limit \maths{\mathcal{V}/\mu\ll1}. An analytical treatment of the dynamics of Bogoliubov excitations in weakly interacting dilute atomic Bose gases subjected to weakly perturbing, spatially correlated disordered potentials was developed in Refs.~\cite{Gaul2011, Gaul2010}. In the present paper, we build upon these works to solve our 1D disordered Bogoliubov-de Gennes problem \eqref{Eq:BdGEquations}.

Rewriting the solutions \maths{\hat{\rho}_{1}(\zeta,\tau)/\sqrt{\rho_{0}(\zeta)}} and \maths{2\,i\,\sqrt{\rho_{0}(\zeta)}\;\hat{\varphi}_{1}(\zeta,\tau)} of Eqs.~\eqref{Eq:BdGEquations} as
\begin{subequations}
\label{Eq:TrickyReformulationDensityPhase}
\begin{align}
\notag
&\left.
\begin{bmatrix}
\hat{\rho}_{1}(\zeta,\tau)/\sqrt{\rho_{0}(\zeta)} \\
2\,i\,\sqrt{\rho_{0}(\zeta)}\;\hat{\varphi}_{1}(\zeta,\tau)
\end{bmatrix}
\right. \\
\label{Eq:TrickyReformulationDensityPhase-a}
&\left.{\quad}=\bigg[\frac{\bar{\rho}_{0}}{\rho_{0}(\zeta)}\bigg]^{\pm\frac{1}{2}}
\begin{bmatrix}
\hat{\rho}_{1}(\zeta,\tau)/\sqrt{\bar{\rho}_{0}} \\
2\,i\,\sqrt{\bar{\rho}_{0}}\;\hat{\varphi}_{1}(\zeta,\tau)
\end{bmatrix}
\right. \\
\label{Eq:TrickyReformulationDensityPhase-b}
&\left.{\quad}=\bigg[\frac{\bar{\rho}_{0}}{\rho_{0}(\zeta)}\bigg]^{\pm\frac{1}{2}}\int\frac{dk}{2\pi}
\begin{bmatrix}
\hat{\rho}_{1}(k,\tau)/\sqrt{\bar{\rho}_{0}} \\
2\,i\,\sqrt{\bar{\rho}_{0}}\;\hat{\varphi}_{1}(k,\tau)
\end{bmatrix}
e^{ik\zeta},\right.
\end{align}
\end{subequations}
applying the Bogoliubov transformation \eqref{Eq:BogoliubovRotation} to \maths{\hat{\rho}_{1}(k,\tau)/\sqrt{\bar{\rho}_{0}}} and \maths{2\,i\,\sqrt{\bar{\rho}_{0}}\;\hat{\varphi}_{1}(k,\tau)} in Eq.~\eqref{Eq:TrickyReformulationDensityPhase-b}, and arranging the resulting formula in the form \eqref{Eq:HermitianAntiHermitianDecomposition}, we get the following \eqref{Eq:BogoliubovFluctuation}-like expansion for the Bogoliubov quantum field \maths{\hat{\gamma}(\zeta,\tau)}:
\begin{align}
\notag
\hat{\gamma}(\zeta,\tau)&\left.=\int\frac{dk}{2\pi}\,[u(k,\zeta)\,e^{ik\zeta}\,\hat{a}(k,\tau)\right. \\
\label{Eq:BogoliubovFluctuationDisorder}
&\left.\hphantom{=}+v(k,\zeta)\,e^{-ik\zeta}\,\hat{a}^{\dag}(k,\tau)].\right.
\end{align}
In this equation, the \maths{\rho_{0}(\zeta)}-dependent weights \maths{u(k,\zeta)} and \maths{v(k,\zeta)} are real and even functions of \maths{k} given by
\begin{equation}
\label{Eq:BogoliubovAmplitudesDisorder}
u(k,\zeta)\pm v(k,\zeta)=(u_{k}\pm v_{k})\,\bigg[\frac{\bar{\rho}_{0}}{\rho_{0}(\zeta)}\bigg]^{\pm\frac{1}{2}}.
\end{equation}
By construction, these disorder-modified Bogoliubov amplitudes comply with the usual \maths{\eta}-orthogonality relation [consistently with Eq.~\eqref{Eq:CommutationRelationBogoliubovOperators}] and with the orthogonality with respect to the deformed classical state \cite{Gaul2011, Gaul2010}:
\begin{align}
\label{Eq:BiorthogonalityRelation}
\int d\zeta~\Psi_{1}^{\dag}(k,\zeta)~\eta~\Psi_{1}^{\vphantom{\dag}}(k',\zeta)&=2\pi\,\delta(k-k'), \\
\label{Eq:OrthogonalityRelationBEC}
\int d\zeta~\Psi_{0}^{\dag}(\zeta,\tau)\,\Psi_{1}^{\vphantom{\dag}}(k,\zeta)&=0,
\end{align}
where \maths{\Psi_{0}(\zeta,\tau)=e^{i\varphi_{0}(\tau)}\sqrt{\rho_{0}(\zeta)}\;{}^{\mathrm{t}}[1~1]}, \maths{\Psi_{1}(k,\zeta)={}^{\mathrm{t}}[u(k,\zeta)~v(k,\zeta)]\,e^{ik\zeta}} is the Bogoliubov wavefunction, and \maths{\eta=\mathrm{diag}(1,-1)} is the Bogoliubov metric. Equation \eqref{Eq:BiorthogonalityRelation} is nothing but the generalization of Eq.~\eqref{Eq:BogoliubovNormalization} when \maths{\rho_{0}=\rho_{0}(\zeta)}. In the weak-disorder limit \maths{\mathcal{V}/\mu\ll1}, we recall that \maths{\rho_{0}(\zeta)} slightly deviates from its unperturbed value \maths{\bar{\rho}_{0}} according to Eq.~\eqref{Eq:ClassicalFluctuationsDensity}. In this case, \maths{u(k,\zeta)} and \maths{v(k,\zeta)} are not too far from their respective disorder-free counterparts \maths{u_{k}} and \maths{v_{k}}. Indeed, at the first order in \maths{|\delta\rho_{0}(\zeta)|/\bar{\rho}_{0}\ll1}, one readily verifies from Eq.~\eqref{Eq:BogoliubovAmplitudesDisorder} that
\begin{equation}
\label{Eq:BogoliubovAmplitudesDisorderApprox}
u(k,\zeta)\pm v(k,\zeta)=(u_{k}\pm v_{k})\,\bigg[1\mp\frac{1}{2}\,\frac{\delta\rho_{0}(\zeta)}{\bar{\rho}_{0}}\bigg],
\end{equation}
which we will use in the following sections to carry on our calculations.

As in Refs.~\cite{Gaul2011, Gaul2010}, we here chose to expand the disorder-dependent Bogoliubov quantum field \maths{\hat{\gamma}(\zeta,\tau)} over the plane-wave eigenbasis of the disorder-free Bogoliubov-de Gennes problem. In this formulation, the Bogoliubov excitations are labeled by a wavenumber \maths{k} independent of the realizations of the disorder. As shown in Eq.~\eqref{Eq:BogoliubovFluctuationDisorder}, they consist in distorted plane waves whose \maths{\zeta}-dependent amplitudes \maths{u(k,\zeta)} and \maths{v(k,\zeta)} are given in Eq.~\eqref{Eq:BogoliubovAmplitudesDisorderApprox} for a weakly perturbing disordered potential. They fulfill the usual \maths{\eta}-orthogonality relation [Eq.~\eqref{Eq:BiorthogonalityRelation}] and, most importantly, decouple from the classical state [Eq.~\eqref{Eq:OrthogonalityRelationBEC}]. At first sight, such a formulation suggests that the plane-wave basis remains an eigenbasis of the disordered Bogoliubov-de Gennes problem \eqref{Eq:BdGEquations}. This is not true in general though, as in the disordered potential a Bogoliubov fluctuation of wavenumber \maths{k} undergoes scattering and, therefore, does not possess a well-defined energy. Correspondingly, it is not clear that the elementary-excitation operator \maths{\hat{a}(k,\tau)} in Eq.~\eqref{Eq:BogoliubovFluctuationDisorder} evolves in a simple harmonic way, as in Eq.~\eqref{Eq:PlaneWaveEvolution}. We treat this important subtlety in the subsequent paragraphs. Most particularly, we discuss the influence of our weak disordered potential on the low-\maths{k} scattering and dispersion properties of the Bogoliubov quantum gas on top of the disordered classical background fluid.

In full generality, a Bogoliubov excitation of wavenumber \maths{k} in the disordered potential \maths{V(\zeta)} does not have a well-defined energy \maths{\epsilon=\tilde{E}_{k}} but rather an energy distribution \maths{S_{k}(\epsilon)} called spectral function \cite{Anderson1997, Muller2011, Trappe2015, Prat2016, Volchkov2018}. Interpreted as an energy density, the spectral function of the disordered Bogoliubov gas is normalized to unity, \maths{\int d\epsilon\,S_{k}(\epsilon)=1}, and in the weak-disorder limit \maths{\mathcal{V}/\mu\ll1}, is given by \cite{Gaul2011, Gaul2010}
\begin{subequations}
\label{Eq:SpectralFunction}
\begin{align}
\label{Eq:SpectralFunction-a}
S_{k}(\epsilon)&=\frac{1}{\pi}\,\frac{\Gamma_{k}/2}{(\epsilon-\tilde{E}_{k})^{2}+(\Gamma_{k}/2)^{2}}, \\
\label{Eq:SpectralFunction-b}
\Gamma_{k}&=-2\,\mathrm{Im}(\Sigma_{k})>0, \\
\label{Eq:SpectralFunction-c}
\tilde{E}_{k}&=E_{k}+\mathrm{Re}(\Sigma_{k}).
\end{align}
\end{subequations}
In these equations, \maths{E_{k}} is the disorder-free Bogoliubov dispersion relation \eqref{Eq:BogoliubovSpectrum} and \maths{\Sigma_{k}=\Sigma(k,\epsilon=E_{k})} is the on-shell self-energy of the disordered Bogoliubov-de Gennes problem \cite{Gaul2011, Gaul2010}. In general, a Bogoliubov excitation of wavenumber \maths{k} thus ends up energy-distributed around a \maths{\tilde{E}_{k}\neq E_{k}} according to the Lorentzian law \eqref{Eq:SpectralFunction-a}. Due to scattering on the random potential \maths{V(\zeta)}, such a quasiparticle correspondingly possesses a finite lifetime \maths{\tau_{k}=\hbar/\Gamma_{k}} provided by the spectral width \maths{\Gamma_{k}}. Of course, \maths{\Sigma(k,\epsilon)=0} in the absence of disorder. In this case, \maths{S_{k}(\epsilon)=\delta(\epsilon-E_{k})} and one recovers that an elementary excitation of wavenumber \maths{k} possesses a single energy \maths{E_{k}}.

In the following, we will specifically focus on the small-wavenumber, \maths{|k|\,\xi\ll1}, regime. In this limit, let us compare the spectral width \maths{\Gamma_{k}} to the unperturbed Bogoliubov dispersion relation \maths{E_{k}}. By introducing the scattering mean free path \maths{\ell_{k}=|\partial E_{k}/\partial(\hbar\,k)|\,\tau_{k}} \cite{Gaul2010, Gaul2011}, where \maths{\partial E_{k}/\partial(\hbar\,k)} is the unperturbed Bogoliubov group velocity, we find
\begin{equation}
\label{Eq:NormalizedLorentzianWidth}
\frac{\Gamma_{k}}{E_{k}}=\frac{|\partial E_{k}/\partial(\hbar\,k)|}{E_{k}/(\hbar\,|k|)}\,\frac{1}{|k|\,\ell_{k}}.
\end{equation}
When \maths{|k|\,\xi\ll1}, the first ratio in the right-hand side tends to unity and the second one is found to be proportional to \maths{(\mathcal{V}/\mu)^{2}\,|k|\,\xi} \cite{Gaul2010, Gaul2011} when truncating the \maths{\mathcal{V}/\mu\ll1} power expansion of \maths{\mathrm{Im}(\Sigma_{k})} at the second order (Born approximation):
\begin{equation}
\label{Eq:NormalizedLorentzianWidthLimit}
\frac{\Gamma_{k}}{E_{k}}\simeq\frac{1}{|k|\,\ell_{k}}\propto\bigg(\frac{\mathcal{V}}{\mu}\bigg)^{2}\,|k|\,\xi,\quad|k|\,\xi\ll1.
\end{equation}
Thus, \maths{\Gamma_{k}} is negligible compared to \maths{E_{k}} when \maths{|k|\,\xi} approaches zero. This implies that the spectral function \eqref{Eq:SpectralFunction-a} can be approximated by a Dirac distribution centered at \maths{\epsilon=\tilde{E}_{k}}:
\begin{equation}
\label{Eq:SpectralFunctionPhonons}
S_{k}(\epsilon)\simeq\delta(\epsilon-\tilde{E}_{k}),\quad|k|\,\xi\ll1.
\end{equation}
As a result, when \maths{|k|\,\xi\ll1}, a Bogoliubov excitation of wavenumber \maths{k} keeps possessing a well-defined energy \maths{\tilde{E}_{k}}, given by the bare Bogoliubov dispersion relation \maths{E_{k}} shifted by the real part of \maths{\Sigma_{k}} [see Eq.~\eqref{Eq:SpectralFunction-c}]. Correspondingly, its annihilation operator \maths{\hat{a}(k,\tau)} follows the harmonic evolution law \eqref{Eq:PlaneWaveEvolution} upon substitution of \maths{E_{k}} by \maths{\tilde{E}_{k}} for \maths{|k|\,\xi\ll1}:
\begin{equation}
\label{Eq:PlaneWaveEvolutionDisorder}
\hat{a}(k,\tau)\simeq e^{-i\tilde{E}_{k}\tau/\hbar}\,\hat{a}(k,0),\quad|k|\,\xi\ll1.
\end{equation}
The energy \maths{\tilde{E}_{k}} for \maths{|k|\,\xi\ll1} is the dispersion relation of the small-wavenumber Bogoliubov fluctuations in the weak disordered potential \maths{V(\zeta)}.

\begin{figure}[t!]
\includegraphics[width=\linewidth]{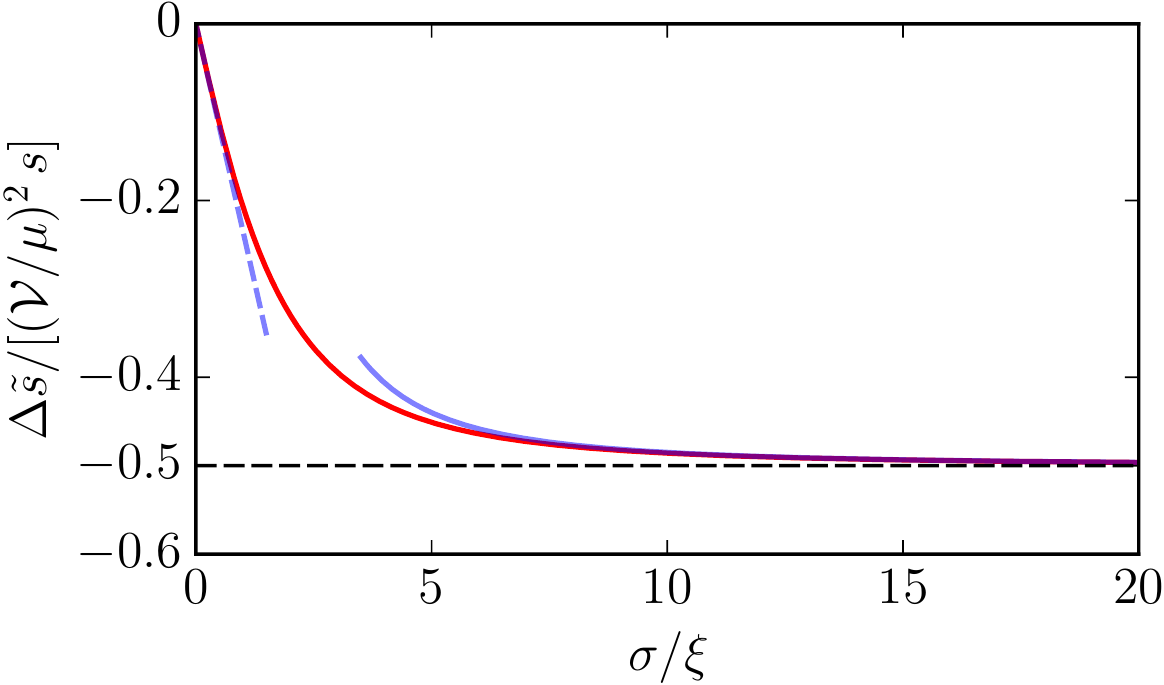}
\caption{(Color online) Red curve: Normalized disorder-induced correction to the Bogoliubov speed of sound as a function of \maths{\sigma/\xi}, as given in Eq.~\eqref{Eq:SoundSpeedRelativeShiftResult}. Blue curves: Asympotic behaviors when \maths{\sigma/\xi\ll1} (dashed curve) and when \maths{\sigma/\xi\gg1} (solid curve), as given in Eq.~\eqref{Eq:SoundSpeedRelativeShiftAsymptoticResult}. The horizontal dashed line indicates that \maths{\Delta\tilde{s}/[(\mathcal{V}/\mu)^{2}\,s]=-1/2} when \maths{\sigma/\xi=\infty} \cite{Gaul2011, Gaul2010}, independently of the original shape of the disorder's correlation function \maths{C(\zeta)}, which in this case equals \maths{1}.}
\label{Fig:SoundVelocity}
\end{figure}

Within the Born approximation for Re(\maths{\Sigma_{k}}), \maths{\tilde{E}_{k}} is given by \cite{Gaul2010, Gaul2011}
\begin{subequations}
\label{Eq:PhononDispersionRelation}
\begin{align}
\label{Eq:PhononDispersionRelation-a}
\tilde{E}_{k}&\simeq\tilde{s}\,\hbar\,|k|=(s+\Delta\tilde{s})\,\hbar\,|k|, \\
\label{Eq:PhononDispersionRelation-b}
\frac{\Delta\tilde{s}}{s}&=-\frac{1}{2}\,\bigg(\frac{\mathcal{V}}{\mu}\bigg)^{2}\int\frac{dk}{2\pi}\,\frac{C(k)}{(k^{2}\,\xi^{2}/2+1)^{3}}.
\end{align}
\end{subequations}
Equation \eqref{Eq:PhononDispersionRelation-a} shows that the dispersion relation \maths{\tilde{E}_{k}} of the small-\maths{k} Bogoliubov fluctuations is phononlike, as in the absence of disorder (\maths{E_{k}\simeq s\,\hbar\,|k|}) but with a disorder-modified speed of sound \maths{\tilde{s}=s+\Delta\tilde{s}}. The sound-velocity relative correction \eqref{Eq:PhononDispersionRelation-b} turns out to be negative, which is a peculiarity of 1D geometries \cite{Gaul2010, Gaul2011}. In Eq.~\eqref{Eq:PhononDispersionRelation-b}, \maths{C(k)=\int d\zeta\,C(\zeta)\,e^{-ik\zeta}} is the space Fourier transform of \maths{C(\zeta)} given in Eqs.~\eqref{Eq:DisorderCorrelations}. Straightforward calculations yield
\begin{align}
\notag
\frac{\Delta\tilde{s}}{s}&\left.=-\frac{3}{16}\,\sqrt{\frac{\pi}{2}}\,\bigg(\frac{\mathcal{V}}{\mu}\bigg)^{2}\,\frac{\sigma}{\xi}\right. \\
\notag
&\left.\hphantom{=}\times\bigg[e^{(\sigma^{2}/\xi^{2})/2}\,\bigg(1-\frac{2}{3}\,\frac{\sigma^{2}}{\xi^{2}}+\frac{1}{3}\,\frac{\sigma^{4}}{\xi^{4}}\bigg)\,\mathrm{erfc}\bigg(\frac{1}{\sqrt{2}}\,\frac{\sigma}{\xi}\bigg)\right. \\
\label{Eq:SoundSpeedRelativeShiftResult}
&\left.\hphantom{=}+\sqrt{\frac{2}{\pi}}\,\frac{\sigma}{\xi}\,\bigg(1-\frac{1}{3}\,\frac{\sigma^{2}}{\xi^{2}}\bigg)\bigg].\right.
\end{align}
In Fig.~\ref{Fig:SoundVelocity}, we plot the corresponding \maths{\Delta\tilde{s}/[(\mathcal{V}/\mu)^{2}\,s]} as a function of \maths{\sigma/\xi}.

Its asymptotic behaviors are also displayed in Fig.~\ref{Fig:SoundVelocity} and they are given by
\begin{equation}
\label{Eq:SoundSpeedRelativeShiftAsymptoticResult}
\frac{\Delta\tilde{s}}{s}\simeq
\begin{dcases}
-\frac{3}{16}\,\sqrt{\frac{\pi}{2}}\,\bigg(\frac{\mathcal{V}}{\mu}\bigg)^{2}\,\frac{\sigma}{\xi}, & \frac{\sigma}{\xi}\ll1, \\
-\frac{1}{2}\,\bigg(\frac{\mathcal{V}}{\mu}\bigg)^{2}\,\bigg(1-\frac{3}{\sigma^{2}/\xi^{2}}\bigg), & \frac{\sigma}{\xi}\gg1.
\end{dcases}
\end{equation}
They are here derived from Eq.~\eqref{Eq:SoundSpeedRelativeShiftResult} but may also be directly obtained from Eq.~\eqref{Eq:PhononDispersionRelation-b}: The limit \maths{\sigma/\xi\ll1} corresponds to an almost uncorrelated disordered potential [\maths{C(\zeta)\simeq\sqrt{\pi}\,\sigma\,\delta(\zeta)} and then \maths{C(k)\simeq\sqrt{\pi}\,\sigma} in Eq.~\eqref{Eq:PhononDispersionRelation-b}], while the limit \maths{\sigma/\xi\gg1} corresponds to a disordered potential with a slowly decaying, quasiparabolic correlation function \{\maths{C(\zeta)\simeq1-\zeta^{2}/\sigma^{2}} and then \maths{C(k)\simeq2\pi\,[1+\partial^{2}/\partial(k\,\sigma)^{2}]\,\delta(k)} in Eq.~\eqref{Eq:PhononDispersionRelation-b}\}. The fact that \maths{\Delta\tilde{s}\propto\mathcal{V}^{2}\,\sigma} when \maths{\sigma/\xi\ll1} is expected since \maths{\mathcal{V}^{2}\,\sigma} is the only combination of \maths{\mathcal{V}} and \maths{\sigma} that an uncorrelated disordered potential can provide: \maths{\overline{V(\zeta)\,V(0)}\propto\mathcal{V}^{2}\,\sigma}. In the opposite limit \maths{\sigma/\xi=\infty}, the result \maths{\Delta\tilde{s}\propto\mathcal{V}^{2}} can be qualitatively understood from a local-density approximation (LDA) where the Bogoliubov wave (of wavelength \maths{\sim\xi}) perceives a locally homogeneous background (of spatial extent \maths{\sim\sigma\gg\xi}). Within this framework, we can define the correction to \maths{s} as \maths{\Delta\tilde{s}_{\mathrm{LDA}}=\overline{s(\zeta)}-s}, where the local sound velocity \maths{s(\zeta)=[\mu(\zeta)/m]^{1/2}} is governed by the local chemical potential \maths{\mu(\zeta)=\mu-V(\zeta)}. We first perform a \maths{\mathcal{V}/\mu\ll1} power expansion at the second order and then carry out the disorder average using \maths{\overline{V(\zeta)\,V(0)}=\mathcal{V}^{2}}, which eventually yields \maths{\Delta\tilde{s}_{\mathrm{LDA}}\propto\mathcal{V}^{2}}.

To summarize, when \maths{|k|\,\xi\ll1}, the spectral function is \maths{\delta}-peaked and the disordered system exhibits a well-defined dispersion relation: It is of phonon type with a specific disorder-renormalized sound velocity. When \maths{|k|\,\xi\gg1} instead, the unperturbed dispersion relation is parabolic and the spectral function becomes broad (see, e.g., Ref.~\cite{Kuhn2007}): No dispersion relation may be identified. This issue could be circumvented by starting from an energy, \maths{\epsilon}, expansion---instead of a wavenumber, \maths{k}, expansion---of the Bogoliubov quantum field \eqref{Eq:BogoliubovFluctuationDisorder}. This was done in, e.g., Refs.~\cite{Larre2012, Larre2013} to describe Bogoliubov quantum fluctuations around nonuniform stationary background patterns in the context of acoustic Hawking radiation. Nevertheless, as we will focus in the following on the phonon regime, the descriptions we established up to here will be sufficient to carry on our calculations.

\section{Quantum quench}
\label{Sec:QuantumQuench}

In Sec.~\ref{Sec:QuantumDensityPhaseBogoliubovTheory}, we treated the quantum fluctuations of the beam \maths{\boldsymbol{1}} in the fiber. Here, we address the effect of the \maths{z=0} and \maths{z=L} interfaces with free space, where \maths{\boldsymbol{1}} evolves independently of \maths{\boldsymbol{2}}. As shown below, these interfaces effectively induce a disorder and interaction quench for the quantum fluid of light.

Upon crossing the entrance of the fiber at \maths{z=0}, the optical nonlinearity is abruptly switched on. The vacuum of the quantum fluctuations of the beam \maths{\boldsymbol{1}} then gets nonadiabatically modified: From free space, it suddenly becomes the Bogoliubov vacuum, and \maths{\boldsymbol{1}} gets in turn projected away from equilibrium. Since \maths{z} plays the role of time, this directly simulates a steplike quench of the quantum fluid of light at \maths{\tau=0/v_{1}=0}. This quench involves the disordered and interaction potentials in Eq.~\eqref{Eq:QuantumNLSFormalismBis-a} since both originate from the optical nonlinearity [see Eqs.~\eqref{Eq:ExternalPotential} and \eqref{Eq:InteractionConstant}]. The propagation distance \maths{z>0} across the fiber simulates the time \maths{\tau=z/v_{1}>0} elapsed after the occurrence of the quench. Thus, when measuring the statistical properties of the light exiting the fiber at \maths{z=L} (see Fig.~\ref{Fig:Setup}), one also gains insight into the nonequilibrium features of the quenched quantum fluid of light at the time \maths{\tau=L/v_{1}=T}.

In Sec.~\ref{SubSec:QuantumOpticalFieldInFreeSpace}, we derive an expression for \maths{\boldsymbol{1}}'s quantum optical field in free space (\maths{z<0} or \maths{z>L}, i.e., \maths{\tau<0} or \maths{\tau>T}). In Sec.~\ref{SubSec:InputOutputRelations} then, we establish the input-output relations connecting the \maths{z<0} and \maths{z>L} regions given \maths{\boldsymbol{1}}'s quantum optical field in the fiber (\maths{0<z<L}, i.e., \maths{0<\tau<T}).

\subsection{Quantum optical field in free space}
\label{SubSec:QuantumOpticalFieldInFreeSpace}

In free space (\maths{z<0} or \maths{z>L}), the beam of quasimonochromatic light \maths{\boldsymbol{1}} is assumed to have a wide top-hat spatial profile in the \maths{x} and \maths{y} directions and to not suffer from attenuation along the \maths{z} axis. In such a configuration, its electric field \maths{\hat{E}_{1}(\mathbf{r},t)} quantum-fluctuates around a classical monochromatic plane wave with angular frequency \maths{\omega_{1}} and wavevector \maths{(\omega_{1}/c_{0})\,\hat{\mathbf{z}}}:
\begin{equation}
\label{Eq:ElectricFieldFreeSpace}
\hat{E}_{1}(\mathbf{r},t)=[\mathcal{E}_{1}+\delta\hat{\mathcal{E}}_{1}(\mathbf{r},t)]\,e^{i[(\omega_{1}/c_{0})z-\omega_{1}t]},
\end{equation}
where \maths{\delta\hat{\mathcal{E}}_{1}(\mathbf{r},t)} is a small and slowly varying quantum departure from the uniform and static classical envelope \maths{\mathcal{E}_{1}}.

Making use of the well-known plane-wave quantization of the electric field in free space \cite{CohenTannoudji2017}, we can express the projection \maths{\hat{A}_{1}^{\vphantom{\ast}}(z,t)=\int dx\,dy\,F_{1}^{\ast}(x,y)\,[\mathcal{E}_{1}^{\vphantom{\ast}}+\delta\hat{\mathcal{E}}_{1}^{\vphantom{\ast}}(\mathbf{r},t)]} of this envelope onto the transverse modal function \maths{F_{1}(x,y)} of the fiber at \maths{\omega_{1}}. It consists in the sum of a homogeneous and stationary classical field and of a weak quantum fluctuation that may be cast in the following form \cite{Larre2016}:
\begin{align}
\notag
\hat{A}_{1}(z,t)&\left.=\bigg(\frac{2\,\hbar\,\omega_{1}}{\varepsilon_{0}}\bigg)^{\frac{1}{2}}\,e^{i\varphi_{0}}\right. \\
\label{Eq:LongitudinalAmplitudeFreeSpace-a}
&\left.\hphantom{=}\times\bigg[\sqrt{\rho_{0}}+\frac{1}{\sqrt{c_{0}}}\int\frac{d\Delta}{2\pi}\,e^{-i\Delta t}\,\delta\hat{\alpha}(\Delta,z)\bigg].\right.
\end{align}
In this equation, valid for \maths{z<0} or \maths{z>L}, the classical linear density \maths{\rho_{0}} and the classical phase \maths{\varphi_{0}} are piecewise constant: \maths{(\rho_{0},\varphi_{0})=(\rho_{\mathrm{in}},\varphi_{\mathrm{in}})} for \maths{z<0} and \maths{(\rho_{0},\varphi_{0})=(\rho_{\mathrm{out}},\varphi_{\mathrm{out}})} for \maths{z>L}. In the quantum-fluctuation term, the integral is taken over the detunings \maths{\Delta=\omega_{\mathbf{q}}-\omega_{1}} from the carrier angular frequency \maths{\omega_{1}}, with \maths{\omega_{\mathbf{q}}=c_{0}\,|\mathbf{q}|=c_{0}\,(q_{x}^{2}+q_{y}^{2}+q_{z}^{2})^{1/2}} the photon dispersion relation in free space. Finally, the operator \maths{\delta\hat{\alpha}(\Delta,z)} derives from the photon annihilation operator in free space \maths{\hat{\alpha}(\mathbf{q})} as \cite{Larre2016}
\begin{align}
\notag
\delta\hat{\alpha}(\Delta,z)&\left.=\frac{i}{\sqrt{c_{0}}}\int\frac{dq_{x}\,dq_{y}}{(2\pi)^{2}}\,F_{1}^{\ast}(q_{x},q_{y})\,e^{i\delta q_{z}(q_{x},q_{y},\Delta)}\right. \\
\label{Eq:LongitudinalAmplitudeFreeSpace-b}
&\left.\hphantom{=}\times\hat{\alpha}\bigg[q_{x},q_{y},\frac{\omega_{1}}{c_{0}}+\delta q_{z}(q_{x},q_{y},\Delta)\bigg],\right.
\end{align}
where \maths{F_{1}(q_{x},q_{y})=\int dx\,dy\,F_{1}(x,y)\,e^{-i(q_{x}x+q_{y}y)}} is the Fourier transform of \maths{F_{1}(x,y)} and \maths{\delta q_{z}(q_{x},q_{y},\Delta)=-(q_{x}^{2}+q_{y}^{2})/[2\,(\omega_{1}/c_{0})]+\Delta/c_{0}} is \maths{q_{z}-\omega_{1}/c_{0}} upon linearization of \maths{\omega_{\mathbf{q}}} around \maths{\mathbf{q}=(0,0,\omega_{1}/c_{0})}. Since \maths{[\hat{\alpha}(\mathbf{q}),\hat{\alpha}^{\dag}(\mathbf{q}')]=(2\pi)^{3}\,\delta(\mathbf{q}-\mathbf{q}')} \cite{CohenTannoudji2017}, the \maths{\delta\hat{\alpha}(\Delta,z)}'s satisfy the following equal-\maths{z} commutation relation:
\begin{equation}
\label{Eq:CommutationRelationDeltaAlpha}
[\delta\hat{\alpha}(\Delta,z),\delta\hat{\alpha}^{\dag}(\Delta',z)]=2\pi\,\delta(\Delta-\Delta').
\end{equation}

To facilitate the matching of the fields at the entrance (\maths{z=0}) and the exit (\maths{z=L}) of the fiber (Sec.~\ref{SubSec:InputOutputRelations}), we are now going to insert the free-space formulas \eqref{Eq:LongitudinalAmplitudeFreeSpace-a}--\eqref{Eq:CommutationRelationDeltaAlpha} into the \maths{z\longleftrightarrow t} mapping used to describe the system's dynamics in the fiber.

For this purpose, we first reintroduce the ``quantum-fluid variables'' \eqref{Eq:TimeVariable} and \eqref{Eq:SpaceVariable}, with here \maths{\tau<0} or \maths{\tau>T}. We then define the free-space counterpart
\begin{subequations}
\label{Eq:MatterFieldFreeSpace}
\begin{align}
\label{Eq:MatterFieldFreeSpace-a}
\hat{\Psi}(\zeta,\tau)&=\bigg(\frac{\mathscr{C}_{0}}{\hbar\,c_{0}}\bigg)^{\frac{1}{2}}\,\hat{A}_{1}\bigg(v_{1}\,\tau,\frac{\zeta}{v_{1}}+\tau\bigg), \\
\label{Eq:MatterFieldFreeSpace-b}
\mathscr{C}_{0}&=\frac{1}{2}\,\frac{c_{0}\,\varepsilon_{0}}{\omega_{1}},
\end{align}
\end{subequations}
of the in-fiber quantum field \eqref{Eq:MatterField}, \eqref{Eq:QuantumFormalism-c}. In Eq.~\eqref{Eq:MatterFieldFreeSpace-a}, the free-space capacitance \maths{\mathscr{C}_{0}} given in Eq.~\eqref{Eq:MatterFieldFreeSpace-b} is nothing but the in-fiber one \eqref{Eq:QuantumFormalism-c} with \maths{(n_{\mathrm{L}})_{1}=1}, and the free-space speed of light \maths{c_{0}} replaces the in-fiber group velocity \maths{v_{1}}. Inserting Eq.~\eqref{Eq:LongitudinalAmplitudeFreeSpace-a} into Eqs.~\eqref{Eq:MatterFieldFreeSpace} and performing the change of variables \maths{k=-\Delta/v_{1}} in the integral over the detunings \maths{\Delta}, we eventually write \maths{\hat{\Psi}(\zeta,\tau)} in the form of a classical contribution corrected by plane-wave quantum modes with wavenumbers \maths{k} along the \maths{\zeta} axis:
\begin{equation}
\label{Eq:MatterFieldFreeSpaceBis-a}
{\hat{\Psi}(\zeta,\tau)=e^{i\varphi_{0}}\,\bigg[\sqrt{\rho_{0}}+\bigg(\frac{v_{1}}{c_{0}}\bigg)^{\frac{1}{2}}\int\frac{dk}{2\pi}\,e^{ik\zeta}\,\hat{a}(k,\tau)\bigg].}
\end{equation}
In the quantum term of Eq.~\eqref{Eq:MatterFieldFreeSpaceBis-a}, the velocity ratio \maths{v_{1}/c_{0}} originates from the fact that we use the same definition for \maths{\tau} and \maths{\zeta} irrespective of whether one is outside or inside the fiber \cite{NoteCoordinates}. The \maths{\hat{a}(k,\tau)}'s are defined in terms of the \maths{\delta\hat{\alpha}(\Delta,z)}'s as
\begin{equation}
\label{Eq:MatterFieldFreeSpaceBis-b}
\hat{a}(k,\tau)=\sqrt{v_{1}}\,e^{ikv_{1}\tau}\,\delta\hat{\alpha}(-v_{1}\,k,v_{1}\,\tau),
\end{equation}
and due to Eq.~\eqref{Eq:CommutationRelationDeltaAlpha}, they satisfy the following equal-\maths{\tau} commutation relation:
\begin{equation}
\label{Eq:CommutationRelationA}
[\hat{a}(k,\tau),\hat{a}^{\dag}(k',\tau)]=2\pi\,\delta(k-k').
\end{equation}
In Eq.~\eqref{Eq:MatterFieldFreeSpaceBis-a}, the quantum field
\begin{equation}
\label{Eq:MatterFieldFreeSpaceBis-c}
\hat{\gamma}(\zeta,\tau)=\bigg(\frac{v_{1}}{c_{0}}\bigg)^{\frac{1}{2}}\int\frac{dk}{2\pi}\,e^{ik\zeta}\,\hat{a}(k,\tau)
\end{equation}
is by construction small compared to the c-number \maths{\sqrt{\rho_{0}}}. As a result, we can resum the first-order expansion \eqref{Eq:MatterFieldFreeSpaceBis-a} in the form \eqref{Eq:MadelungRepresentation}, \eqref{Eq:QuantumFluctuationsDensity}, \eqref{Eq:QuantumFluctuationsPhase}, \eqref{Eq:HermitianAntiHermitianDecomposition}:
\begin{equation}
\label{Eq:MatterFieldFreeSpaceTer-a}
\hat{\Psi}(\zeta,\tau)=e^{i\hat{\varphi}(\zeta,\tau)}\sqrt{\hat{\rho}(\zeta,\tau)},
\end{equation}
where the quantum linear density \maths{\hat{\rho}(\zeta,\tau)} and the quantum phase \maths{\hat{\varphi}(\zeta,\tau)} are expanded as
\begin{align}
\label{Eq:MatterFieldFreeSpaceTer-b}
\hat{\rho}(\zeta,\tau)&=\rho_{0}+\hat{\rho}_{1}(\zeta,\tau), \\
\label{Eq:MatterFieldFreeSpaceTer-c}
\hat{\varphi}(\zeta,\tau)&=\varphi_{0}+\hat{\varphi}_{1}(\zeta,\tau),
\end{align}
the quantum contributions of which are symmetrically expressed as
\begin{equation}
\label{Eq:MatterFieldFreeSpaceTer-d}
\begin{bmatrix}
\hat{\rho}_{1}(\zeta,\tau)/\sqrt{\rho_{0}} \\
2\,i\,\sqrt{\rho_{0}}\;\hat{\varphi}_{1}(\zeta,\tau)
\end{bmatrix}
=\hat{\gamma}(\zeta,\tau)\pm\hat{\gamma}^{\dag}(\zeta,\tau).
\end{equation}

Equations \eqref{Eq:CommutationRelationA}--\eqref{Eq:MatterFieldFreeSpaceTer-d} constitute the reformulation of \maths{\boldsymbol{1}}'s quantum optical field in free space (\maths{z<0} or \maths{z>L}) within the \maths{z\longleftrightarrow t} language used to describe the system's dynamics in the fiber (\maths{0<z<L}). This facilitates the matching of the fields at \maths{z=0} and \maths{z=L}, as detailed in the next section.

\subsection{Input-output relations}
\label{SubSec:InputOutputRelations}

As in Refs.~\cite{Larre2015, Larre2016, Larre2017}, we assume that the entrance (\maths{z=0}) and the exit (\maths{z=L}) facets of the fiber are treated with an ideal antireflection coating. In such a configuration, all the back-propagating modes originating from light reflection on the \maths{z=0} and \maths{z=L} diopters are suppressed and light transmission across the fiber is perfect. This constrains \maths{\boldsymbol{1}}'s envelope to propagate in the positive-\maths{z} direction and makes the \maths{z\longleftrightarrow t} mapping legitimate. Such an antireflection coating has a characteristic thickness of the order of a few optical wavelengths, then much shorter than any other length scale in the considered problem. Therefore, its effect on light transmission can be described as simple boundary conditions guaranteeing the continuity of the flux of the Poynting vector of the optical beam \maths{\boldsymbol{1}} at both \maths{z=0} and \maths{z=L}.

In mathematical terms, these continuity conditions form the following system:
\begin{subequations}
\label{Eq:ContinuityPoyntingVector}
\begin{align}
\label{Eq:ContinuityPoyntingVector-a}
\int dx\,dy\,\langle\hat{\Pi}_{1}(\mathbf{r},t)\rangle_{t}\bigg|_{z=0^{-}}&=\int dx\,dy\,\langle\hat{\Pi}_{1}(\mathbf{r},t)\rangle_{t}\bigg|_{z=0^{+}}, \\
\label{Eq:ContinuityPoyntingVector-b}
\int dx\,dy\,\langle\hat{\Pi}_{1}(\mathbf{r},t)\rangle_{t}\bigg|_{z=L^{-}}&=\int dx\,dy\,\langle\hat{\Pi}_{1}(\mathbf{r},t)\rangle_{t}\bigg|_{z=L^{+}}.
\end{align}
\end{subequations}
In these equations, \maths{\hat{\Pi}_{1}(\mathbf{r},t)} is the quantum field associated with the \maths{z} component of \maths{\boldsymbol{1}}'s Poynting vector and \maths{\langle{\cdots}\rangle_{t}=(2\pi/\omega_{1})^{-1}\int_{0}^{2\pi/\omega_{1}}dt\,({\cdots})} accounts for the fact that the photodetectors perform an average over at least one time period \maths{2\pi/\omega_{1}} of the carrier. In the fiber (\maths{0<z<L}), the flux \maths{\int dx\,dy\,\langle\hat{\Pi}_{1}(\mathbf{r},t)\rangle_{t}} is expressed as (see Appendix \ref{App:FluxOfThePoyntingVectorOfACrossPhaseModulatedOpticalBeam})
\begin{align}
\notag
&\left.\int dx\,dy\,\langle\hat{\Pi}_{1}(\mathbf{r},t)\rangle_{t}\right. \\
\label{Eq:FluxPoyntingVectorFiber}
&\left.{\quad}=\frac{1}{2}\,c_{0}\,\varepsilon_{0}\,(n_{\mathrm{L}})_{1}\,\mathcal{F}(v_{1}\,t-z)\,\hat{A}_{1}^{\dag}(z,t)\,\hat{A}_{1}^{\vphantom{\dag}}(z,t),\right.
\end{align}
where \maths{\mathcal{F}(v_{1}\,t-z)} is given in Eq.~\eqref{Eq:StructureFactor}. In free space on the other hand (\maths{z<0} or \maths{z>L}), it admits the simple expression
\begin{equation}
\label{Eq:FluxPoyntingVectorFreeSpace}
\int dx\,dy\,\langle\hat{\Pi}_{1}(\mathbf{r},t)\rangle_{t}=\frac{1}{2}\,c_{0}\,\varepsilon_{0}\,\hat{A}_{1}^{\dag}(z,t)\,\hat{A}_{1}^{\vphantom{\dag}}(z,t),
\end{equation}
which is obtained from Eq.~\eqref{Eq:FluxPoyntingVectorFiber} by setting \maths{(n_{\mathrm{L}})_{1}=1} and \maths{\Delta n_{\mathrm{L}}(x,y,\omega_{1})=(\Delta n_{\mathrm{L}})'(\mathbf{r},\omega_{1})=0}. Reformulating Eqs.~\eqref{Eq:ContinuityPoyntingVector} in the \maths{z\longleftrightarrow t} language, we come to
\begin{subequations}
\label{Eq:ContinuityPoyntingVectorBis}
\begin{gather}
\label{Eq:ContinuityPoyntingVectorBis-a}
\rho_{\mathrm{in}}+\hat{\rho}_{1}(\zeta,0^{-})=\frac{v_{1}}{c_{0}}\,\mathcal{F}(\zeta)\,[\rho_{0}(\zeta)+\hat{\rho}_{1}(\zeta,0^{+})], \\
\label{Eq:ContinuityPoyntingVectorBis-b}
\frac{v_{1}}{c_{0}}\,\mathcal{F}(\zeta)\,[\rho_{0}(\zeta)+\hat{\rho}_{1}(\zeta,T^{-})]=\rho_{\mathrm{out}}+\hat{\rho}_{1}(\zeta,T^{+}).
\end{gather}
\end{subequations}

At the classical level, \maths{\hat{\rho}_{1}(\zeta,\tau)=0} and Eqs.~\eqref{Eq:ContinuityPoyntingVectorBis} then reduce to
\begin{equation}
\label{Eq:ContinuityPoyntingVectorClassicalLevel}
\rho_{\mathrm{out}}=\rho_{\mathrm{in}}.
\end{equation}
Using \maths{\hat{\rho}_{1}/\sqrt{\rho_{0}}=\hat{\gamma}+\hat{\gamma}^{\dag}}, we now look after the quantum-fluctuation terms in Eqs.~\eqref{Eq:ContinuityPoyntingVectorBis}. Combining Eqs.~\eqref{Eq:BogoliubovFluctuationDisorder}, \eqref{Eq:BiorthogonalityRelation}, \eqref{Eq:OrthogonalityRelationBEC}, \eqref{Eq:PlaneWaveEvolutionDisorder}, and \eqref{Eq:MatterFieldFreeSpaceBis-c}, we find after straightforward manipulations \cite{Larre2015, Larre2016}
\begin{align}
\notag
\hat{\gamma}_{\mathrm{out}}(\zeta)&\left.=\bigg(\frac{v_{1}}{c_{0}}\bigg)^{\frac{1}{2}}\int\frac{dk}{2\pi}\,[u_{T}(k,\zeta)\,e^{ik\zeta}\,\hat{a}_{\mathrm{in}}(k)\right. \\
\label{Eq:InputOutputRelation}
&\left.\hphantom{=}+v_{T}^{\ast}(k,\zeta)\,e^{-ik\zeta}\,\hat{a}_{\mathrm{in}}^{\dag}(k)],\right.
\end{align}
where \maths{\hat{\gamma}_{\mathrm{out}}(\zeta)=\hat{\gamma}(\zeta,T^{+})}, \maths{\hat{a}_{\mathrm{in}}(k)=\hat{a}(k,0^{-})}, and
\begin{align}
\label{Eq:BogoliubovAmplitudeQuenchU}
u_{T}(k,\zeta)&=u^{2}(k,\zeta)\,e^{-i\tilde{E}_{k}T/\hbar}-v^{2}(k,\zeta)\,e^{i\tilde{E}_{k}T/\hbar}, \\
\label{Eq:BogoliubovAmplitudeQuenchV}
v_{T}(k,\zeta)&=u(k,\zeta)\,v(k,\zeta)\,(e^{-i\tilde{E}_{k}T/\hbar}-e^{i\tilde{E}_{k}T/\hbar}).
\end{align}
The classical and quantum input-output relations \eqref{Eq:ContinuityPoyntingVectorClassicalLevel} and \eqref{Eq:InputOutputRelation} fix the interdependency between the incoming and outgoing light fields given the system's dynamics in the fiber. From them, it is straightforward to extract the quantum coherence properties of the randomly cross-phase modulated beam of light \maths{\boldsymbol{1}} exiting the nonlinear optical fiber, which we analyze in the next section.

Noticeably, the formulation \eqref{Eq:InputOutputRelation} is typical of a small-amplitude quench. Indeed, the quantum field resulting from such a quench can always be Bogoliubov-expanded over the prequench oscillators \maths{\hat{a}_{\mathrm{in}}^{\vphantom{\dag}}(k)} and \maths{\hat{a}_{\mathrm{in}}^{\dag}(-k)} with Bogoliubov-type amplitudes \maths{u_{T}(k,\zeta)} and \maths{v_{T}(k,\zeta)} depending on the nature of the quench and on the time \maths{T} elapsed after its occurrence (see, e.g., Ref.~\cite{Carusotto2010}).

Futhermore, it should be stressed that the phonon limit \maths{|k|\,\xi\ll1} is implicitly considered in the latter equations, precisely because we made use of Eq.~\eqref{Eq:PlaneWaveEvolutionDisorder} to derive them. Correspondingly, the in-fiber Bogoliubov amplitudes \maths{u(k,\zeta)} and \maths{v(k,\zeta)} defined through Eq.~\eqref{Eq:BogoliubovAmplitudesDisorderApprox} must be evaluated in this limit. As one anticipates from Eqs.~\eqref{Eq:BogoliubovAmplitudeQuenchU} and \eqref{Eq:BogoliubovAmplitudeQuenchV}, this is fully true as soon as \maths{T\to\infty}. In this limit indeed, the integral over the wavenumber \maths{k} in Eq.~\eqref{Eq:InputOutputRelation} is dominated by the Bogoliubov modes with \maths{\tilde{E}_{k}\to0}, and so with \maths{|k|\to0} since \maths{\tilde{E}_{k}\propto|k|} as \maths{|k|\to0}. In the proper units, Eqs.~\eqref{Eq:InputOutputRelation}--\eqref{Eq:BogoliubovAmplitudeQuenchV} are actually valid when
\begin{equation}
\label{Eq:LargeT-LowK}
|k|\,\xi\ll\frac{\hbar}{\mu\,T}\ll1,
\end{equation}
to which we restrict ourselves from now on. Coming back to the original coordinates \maths{z} and \maths{t}, this amounts to consider a large optical-fiber length \maths{L=v_{1}\,T} as well as small angular-frequency detunings \maths{\Delta=-v_{1}\,k}, precisely such that \maths{(\xi/v_{1})\,|\Delta|\ll(\hbar\,v_{1}/\mu)\,L^{-1}\ll1}.

\section{Postquench coherence}
\label{Sec:PostquenchCoherence}

Assuming that the electric field measured at \maths{z=0^{-}} (i.e., just before the quench) is a perfect monochromatic plane wave, one has \maths{\langle\hat{E}_{1}(x,y,0^{-},t)\rangle=\mathcal{E}_{1}\,e^{-i\omega_{1}t}}, where \maths{\langle{\cdots}\rangle=\langle\mathrm{vac}|{\cdots}|\mathrm{vac}\rangle} stands for the expectation value in the vacuum state \maths{|\mathrm{vac}\rangle} of \maths{\delta\hat{\mathcal{E}}_{1}(x,y,0^{-},t)}. According to Sec.~\ref{Sec:QuantumQuench}, this can be translated into \maths{\hat{a}_{\mathrm{in}}(k)\,|\mathrm{vac}\rangle=0,\,\forall k}. Therefore, one initially has
\begin{equation}
\label{Eq:InitialCondition1}
\langle\hat{a}_{\mathrm{in}}^{\vphantom{\dag}}(k)\,\hat{a}_{\mathrm{in}}^{\vphantom{\dag}}(k')\rangle=\langle\hat{a}_{\mathrm{in}}^{\dag}(k)\,\hat{a}_{\mathrm{in}}^{\vphantom{\dag}}(k')\rangle=0
\end{equation}
and, making use of the same-\maths{\tau} commutation relation in free space \eqref{Eq:CommutationRelationA},
\begin{equation}
\label{Eq:InitialCondition2}
\langle\hat{a}_{\mathrm{in}}^{\vphantom{\dag}}(k)\,\hat{a}_{\mathrm{in}}^{\dag}(k')\rangle=2\pi\,\delta(k-k').
\end{equation}

In this section, we analyze the consequences of the quench at \maths{\tau=0} through the coherence function
\begin{equation}
\label{Eq:AutocorrelationFunction}
g^{(1)}(\zeta-\zeta')=\overline{\langle\hat{\Psi}^{\dag}(\zeta,T^{+})\,\hat{\Psi}(\zeta',T^{+})\rangle}
\end{equation}
of the field \maths{\hat{\Psi}(\zeta,T^{+}=L^{+}/v_{1})} just exiting the fiber, where light is imaged (see Fig.~\ref{Fig:Setup}). In Eq.~\eqref{Eq:AutocorrelationFunction}, the overbar refers to disorder averaging. Thus defined, \maths{g^{(1)}} only depends on \maths{|\zeta-\zeta'|} since \maths{\rho_{\mathrm{out}}=\mathrm{const}} and \maths{\bar{\rho}_{0}=\mathrm{const}}. Coming back to the original space and time variables \maths{z} and \maths{t}, this means that it only depends on \maths{|t-t'|}. Note that imaging the signal at \maths{z>L} amounts to calculate the \maths{g^{(1)}} function of the field \maths{\hat{\Psi}(\zeta,\tau>T)}. The latter is given in Eqs.~\eqref{Eq:CommutationRelationA}--\eqref{Eq:MatterFieldFreeSpaceTer-d} but can alternatively be obtained from Kirchhoff's diffraction formula for nonmonochromatic waves \cite{Goodman1968}, as sketched in Ref.~\cite{Larre2016}.

\subsection{General formulas}
\label{SubSec:GeneralFormulas}

In the nonlinear optical fiber, the 1D quantum fluid of light is weakly interacting. In this case, the coherence function \eqref{Eq:AutocorrelationFunction} is expressed in terms of the density and the phase quantum fluctuations \maths{\hat{\rho}_{1}(\zeta,T^{+})} and \maths{\hat{\varphi}_{1}(\zeta,T^{+})} of the field \maths{\hat{\Psi}(\zeta,T^{+})} in the following form \cite{Mora2003, Larre2013}:
\begin{align}
\notag
&\left.\ln\!\bigg[\frac{g^{(1)}(\zeta-\zeta')}{\rho_{\mathrm{out}}}\bigg]\right. \\
\notag
&\left.{\quad}=-\frac{1}{8}\,\overline{\bigg\langle\mathopen{:\,}\bigg[\frac{\hat{\rho}_{1}(\zeta,T^{+})}{\rho_{\mathrm{out}}}-\frac{\hat{\rho}_{1}(\zeta',T^{+})}{\rho_{\mathrm{out}}}\bigg]^{2}\mathopen{\,:}\bigg\rangle}\right. \\
\label{Eq:AutocorrelationFunctionBis}
&\left.\hphantom{{\quad}=}-\frac{1}{2}\,\overline{\langle\mathopen{:\,}[\hat{\varphi}_{1}(\zeta,T^{+})-\hat{\varphi}_{1}(\zeta',T^{+})]^{2}\mathopen{\,:}\rangle}.\right.
\end{align}
This formula involves the background density \maths{\rho_{0}(T^{+})=\rho_{\mathrm{out}}} of the outgoing optical beam \maths{\boldsymbol{1}} and the normal ordering \maths{\mathopen{:\,}{\cdots}\mathopen{\,:}} with respect to the Bogoliubov-type quantum field \maths{\hat{\gamma}(\zeta,T^{+})=\hat{\gamma}_{\mathrm{out}}(\zeta)} as a function of which \maths{\hat{\rho}_{1}(\zeta,T^{+})} and \maths{\hat{\varphi}_{1}(\zeta,T^{+})} are defined [see Eq.~\eqref{Eq:MatterFieldFreeSpaceTer-d}]. To obtain Eq.~\eqref{Eq:AutocorrelationFunctionBis}, we proceed in two steps. First, we evaluate the average \maths{\langle{\cdots}\rangle} over the quantum fluctuations of the incoming field. To do so, we use that \maths{\hat{\rho}_{1}(\zeta,T^{+})} is small and that the latter and \maths{\hat{\varphi}_{1}(\zeta,T^{+})} are Gaussianly distributed at the here-considered Bogoliubov level \cite{Mora2003}. Second, we evaluate the average \maths{\overline{{\cdots}}} over the classical fluctuations of the disordered potential. To do so, we take advantage of the fact that the quantum average \maths{\langle\mathopen{:\,}[{\cdots}]^{2}\mathopen{\,:}\rangle} involving the phase fluctuations in Eq.~\eqref{Eq:AutocorrelationFunctionBis} is---due to the normal ordering---as small as the one involving the density fluctuations \cite{Larre2013} (indeed, this correlator looks closely like the two-point correlation function of the velocity field, which is weakly fluctuating). In this case, the approximation \maths{\ln\overline{\exp X}\simeq\overline{X}} holds, which eventually yields Eq.~\eqref{Eq:AutocorrelationFunctionBis}. Note that a Popov approach \cite{Popov1972, Popov1983} for calculating the quantum averages would have yielded the same result \cite{Larre2013}.

Inserting the input-output relations \eqref{Eq:ContinuityPoyntingVectorClassicalLevel} and \eqref{Eq:InputOutputRelation} into Eq.~\eqref{Eq:AutocorrelationFunctionBis} and making use of Eqs.~\eqref{Eq:InitialCondition1} and \eqref{Eq:InitialCondition2}, we obtain, after multiplying by \maths{(c_{0}/v_{1})\,\rho_{\mathrm{in}}\,\xi},
\begin{align}
\notag
&\left.\frac{c_{0}}{v_{1}}\,\rho_{\mathrm{in}}\,\xi\;\ln\!\bigg[\frac{g^{(1)}(\zeta-\zeta')}{\rho_{\mathrm{in}}}\bigg]\right. \\
\label{Eq:AutocorrelationFunctionTer}
&\left.{\quad}=-\int\frac{dk\,\xi}{2\pi}\,\frac{\overline{|v_{T}(k,\zeta)\,e^{ik\zeta}-v_{T}(k,\zeta')\,e^{ik\zeta'}|^{2}}}{2}.\right.
\end{align}
Note that this equation involves the second quench-modified Bogoliubov amplitude, \maths{v_{T}(k,\zeta)}, but not the first one, \maths{u_{T}(k,\zeta)}. This is due to the fact that we assumed all the fluctuation modes of the incident quantum field to be in the vacuum state. If we were in a configuration where \maths{\langle\hat{a}_{\mathrm{in}}^{\dag}(k)\,\hat{a}_{\mathrm{in}}^{\vphantom{\dag}}(k')\rangle\neq0}, the \maths{g^{(1)}} function would present a \maths{u_{T}(k,\zeta)} dependence, as for a weakly interacting dilute atomic Bose gas at thermal equilibrium \cite{Dalfovo1999, Pitaevskii2016}.

Plugging Eq.~\eqref{Eq:BogoliubovAmplitudeQuenchV} supplemented by Eq.~\eqref{Eq:BogoliubovAmplitudesDisorderApprox} into Eq.~\eqref{Eq:AutocorrelationFunctionTer} and making use of Eq.~\eqref{Eq:DensityCorrelations}, we then get
\begin{align}
\notag
&\left.\frac{c_{0}}{v_{1}}\,\rho_{\mathrm{in}}\,\xi\;\ln\!\bigg[\frac{g^{(1)}(\zeta-\zeta')}{\rho_{\mathrm{in}}}\bigg]\right. \\
\label{Eq:AutocorrelationFunctionQuater}
&\left.{\quad}=-I_{T}(X_{1};\zeta-\zeta')-\frac{3}{2}\,\frac{G(0)}{\bar{\rho}_{0}^{2}}\,I_{T}[X_{2}(\zeta-\zeta');\zeta-\zeta'],\right.
\end{align}
where we introduced the short-hand notations
\begin{align}
\label{Eq:X1}
X_{1}&=1, \\
\label{Eq:X2}
X_{2}(\zeta-\zeta')&=\frac{1}{3}\,\bigg[1+2\,\frac{G(\zeta-\zeta')}{G(0)}\bigg].
\end{align}
Here, we made an expansion at the second order in \maths{|\delta\rho_{0}(\zeta)|/\bar{\rho}_{0}\sim\mathcal{V}/\mu\ll1}: The first term in the right-hand side of Eq.~\eqref{Eq:AutocorrelationFunctionQuater} is of the order of \maths{(\mathcal{V}/\mu)^{0}=1} while the second one is of the order of \maths{(\mathcal{V}/\mu)^{2}}. These terms involve the position- and time-dependent integral
\begin{align}
\notag
I_{T}(X;\zeta-\zeta')&\left.=\int\frac{dk\,\xi}{2\pi}\,\frac{\sin^{2}(k\,\zeta_{T}/2)}{k^{2}\,\xi^{2}}\right. \\
\label{Eq:TrapezeFunction}
&\left.\hphantom{=}\times[1-X\cos(k\,|\zeta-\zeta'|)],\right.
\end{align}

In Eq.~\eqref{Eq:TrapezeFunction}, the integrand was evaluated in the large-\maths{T}, small-\maths{k} limit \eqref{Eq:LargeT-LowK}. In this form, the integral simply reduces to a two-step trapezoidal function of \maths{|\zeta-\zeta'|} that is linear up to
\begin{equation}
\label{Eq:LightConeBoundary}
|\zeta-\zeta'|=\zeta_{T}=2\,\tilde{s}\,T=2\,s\,T\,\bigg(1+\frac{\Delta\tilde{s}}{s}\bigg)
\end{equation}
and that stays constant above:
\begin{align}
\notag
&\left.I_{T}(X;\zeta-\zeta')\right. \\
\label{Eq:TrapezeFunctionBis}
&\left.{\quad}=
\begin{dcases}
\frac{X}{4}\,\frac{|\zeta-\zeta'|}{\xi}+\frac{1-X}{4}\,\frac{\zeta_{T}}{\xi}, & |\zeta-\zeta'|<\zeta_{T}, \\
\frac{1}{4}\,\frac{\zeta_{T}}{\xi}, & |\zeta-\zeta'|>\zeta_{T}.
\end{dcases}
\right.
\end{align}
Inserting the explicit expression \eqref{Eq:TrapezeFunctionBis} for \maths{X=X_{1}} [Eq.~\eqref{Eq:X1}] and \maths{X=X_{2}(\zeta-\zeta')} [Eq.~\eqref{Eq:X2}] into Eq.~\eqref{Eq:AutocorrelationFunctionQuater}, we obtain a closed analytical expression for the coherence function \eqref{Eq:AutocorrelationFunction}, as detailed below.

\subsubsection{Case where \texorpdfstring{\maths{|\zeta-\zeta'|<\zeta_{T}}}{Lg}}
\label{SubSubSec:CaseWhere<}

The \maths{g^{(1)}} function depends on \maths{|\zeta-\zeta'|} and is given by
\begin{align}
\notag
&\left.\frac{c_{0}}{v_{1}}\,\rho_{\mathrm{in}}\,\xi\;\ln\!\bigg[\frac{g^{(1)}(\zeta-\zeta')}{\rho_{\mathrm{in}}}\bigg]\right. \\
\notag
&\left.{\quad}=-\frac{1}{2}\,\frac{\tilde{\theta}_{\mathrm{eff}}}{\mu}\,\frac{|\zeta-\zeta'|}{\xi}-\frac{1}{4}\,\frac{G(0)}{\bar{\rho}_{0}^{2}}\,\frac{\zeta_{T}}{\xi}\right. \\
\label{Eq:AutocorrelationFunctionQuinquies}
&\left.\hphantom{{\quad}=}+\frac{1}{4}\,\bigg(1-\frac{|\zeta-\zeta'|}{\zeta_{T}}\bigg)\,\frac{G(\zeta-\zeta')}{\bar{\rho}_{0}^{2}}\,\frac{\zeta_{T}}{\xi}.\right.
\end{align}

Since \maths{\mu} is an energy, the quantity \maths{\tilde{\theta}_{\mathrm{eff}}} may be referred to as a temperature in units of the Boltzmann constant. It weakly deviates from its disorder-free counterpart \maths{\theta_{\mathrm{eff}}=\mu/2} as
\begin{subequations}
\label{Eq:EffectiveTemperature}
\begin{align}
\label{Eq:EffectiveTemperature-a}
\tilde{\theta}_{\mathrm{eff}}&\left.=\theta_{\mathrm{eff}}+\Delta\tilde{\theta}_{\mathrm{eff}},\right. \\
\notag
\frac{\Delta\tilde{\theta}_{\mathrm{eff}}}{\theta_{\mathrm{eff}}}&\left.=\frac{1}{2}\,\frac{G(0)}{\bar{\rho}_{0}^{2}}=\frac{\sqrt{\pi}}{4}\,\bigg(\frac{\mathcal{V}}{\mu}\bigg)^{2}\,\frac{\sigma}{\xi}\,\bigg[e^{\sigma^{2}/\xi^{2}}\,\bigg(1-2\,\frac{\sigma^{2}}{\xi^{2}}\bigg)\right. \\
\label{Eq:EffectiveTemperature-b}
&\left.\hphantom{=\frac{1}{2}\,\frac{G(0)}{\bar{\rho}_{0}^{2}}=}\times\mathrm{erfc}\bigg(\frac{\sigma}{\xi}\bigg)+\frac{2}{\sqrt{\pi}}\,\frac{\sigma}{\xi}\bigg].\right.
\end{align}
\end{subequations}
The last equality in Eqs.~\eqref{Eq:EffectiveTemperature-b} follows from Eqs.~\eqref{Eq:DensityCorrelationsResult}. When \maths{\sigma/\xi\ll1} or \maths{\sigma/\xi\gg1}, the disorder-induced relative correction to \maths{\theta_{\mathrm{eff}}} reduces to
\begin{equation}
\label{Eq:EffectiveTemperatureAsymptoticResult}
\frac{\Delta\tilde{\theta}_{\mathrm{eff}}}{\theta_{\mathrm{eff}}}\simeq
\begin{dcases}
\displaystyle{\frac{\sqrt{\pi}}{4}\,\bigg(\frac{\mathcal{V}}{\mu}\bigg)^{2}\,\frac{\sigma}{\xi}}, & \displaystyle{\frac{\sigma}{\xi}\ll1}, \\
\displaystyle{\frac{1}{2}\,\bigg(\frac{\mathcal{V}}{\mu}\bigg)^{2}\,\bigg(1-\frac{1}{\sigma^{2}/\xi^{2}}\bigg)}, & \displaystyle{\frac{\sigma}{\xi}\gg1}.
\end{dcases}
\end{equation}

From Eq.~\eqref{Eq:AutocorrelationFunctionQuinquies}, one may extract the behavior of the \maths{g^{(1)}} function at very short \maths{|\zeta-\zeta'|}:
\begin{align}
\notag
&\left.\frac{c_{0}}{v_{1}}\,\rho_{\mathrm{in}}\,\xi\;\ln\!\bigg[\frac{g^{(1)}(\zeta-\zeta')}{\rho_{\mathrm{in}}}\bigg]\right. \\
\notag
&\left.{\quad}\simeq-\frac{1}{4}\,\bigg[1+\frac{3}{2}\,\frac{G(0)}{\bar{\rho}_{0}^{2}}\bigg]\,\frac{|\zeta-\zeta'|}{\xi}\right. \\
\label{Eq:AutocorrelationFunctionVeryShortRanges}
&\left.\hphantom{{\quad}\simeq}+\frac{1}{8}\,\frac{\zeta_{T}}{\xi}\,\frac{\partial^{2}(G/\bar{\rho}_{0}^{2})}{\partial|\zeta-\zeta'|^{2}}(0)\,(\zeta-\zeta')^{2}.\right.
\end{align}

\subsubsection{Case where \texorpdfstring{\maths{|\zeta-\zeta'|>\zeta_{T}}}{Lg}}
\label{SubSubSec:CaseWhere>}

The \maths{g^{(1)}} function stays locked to the value it takes at \maths{|\zeta-\zeta'|=\zeta_{T}} [Eq.~\eqref{Eq:AutocorrelationFunctionQuinquies} for \maths{|\zeta-\zeta'|=\zeta_{T}}] and then no longer depends on \maths{|\zeta-\zeta'|}:
\begin{equation}
\label{Eq:LongRangePlateau}
{\frac{c_{0}}{v_{1}}\,\rho_{\mathrm{in}}\,\xi\;\ln\!\bigg[\frac{g^{(1)}(\zeta-\zeta')}{\rho_{\mathrm{in}}}\bigg]=-\frac{1}{4}\,\bigg[1+\frac{3}{2}\,\frac{G(0)}{\bar{\rho}_{0}^{2}}\bigg]\,\frac{\zeta_{T}}{\xi}.}
\end{equation}

According to Eqs.~\eqref{Eq:AutocorrelationFunctionVeryShortRanges} and \eqref{Eq:LongRangePlateau}, the curve's points of abscissas \maths{|\zeta-\zeta'|=0} and \maths{|\zeta-\zeta'|=\zeta_{T}} belong to the straight line of slope \maths{-\frac{1}{4}\,[1+\frac{3}{2}\,G(0)/\bar{\rho}_{0}^{2}]}.

\subsection{Prethermalization in disorder}
\label{SubSec:PrethermalizationInDisorder}

\begin{figure}[t!]
\includegraphics[width=\linewidth]{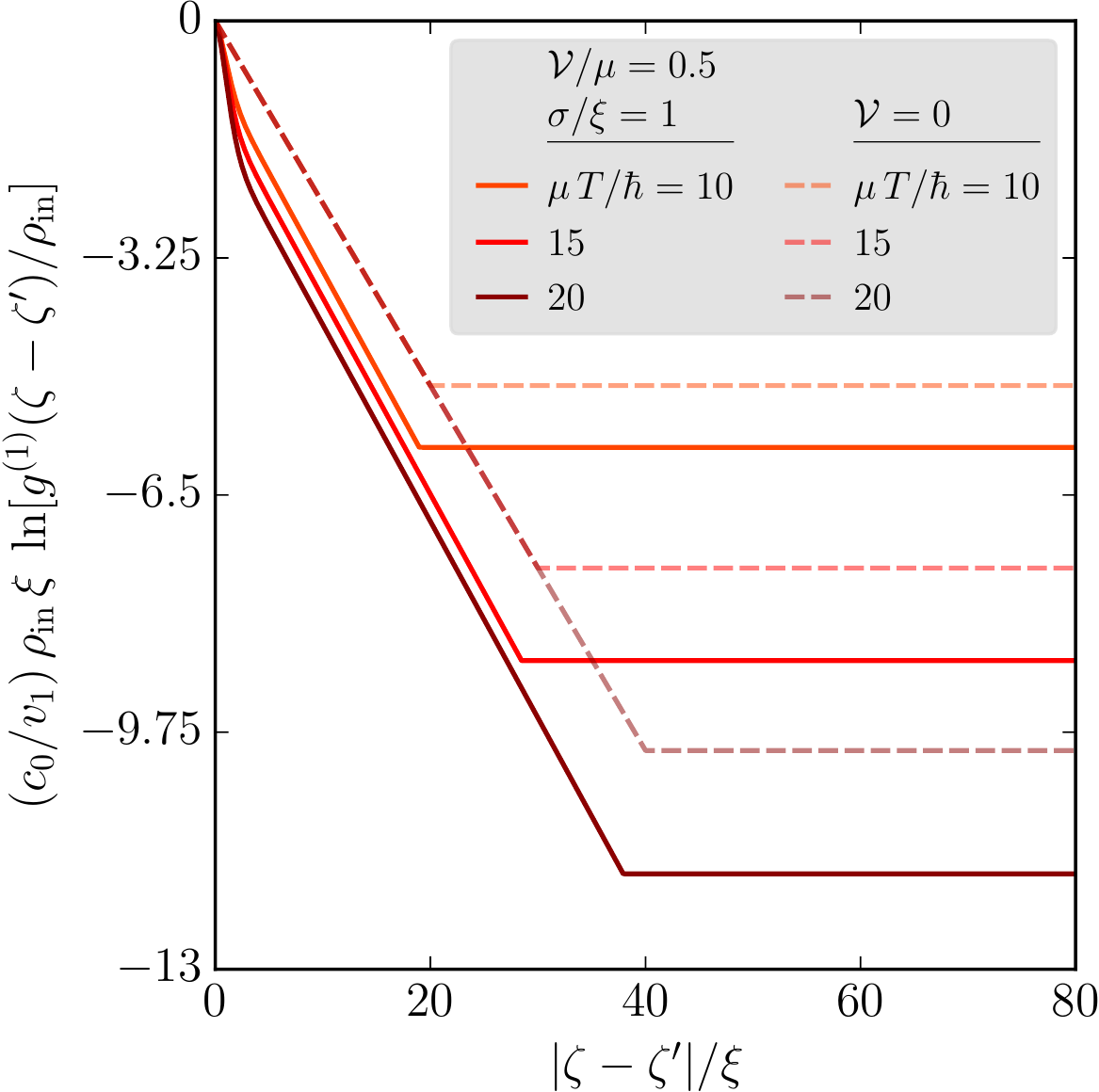}
\caption{(Color online) Solid curves: Normalized coherence function of the disordered 1D quantum fluid of light as a function of \maths{|\zeta-\zeta'|/\xi} for different values of the dimensionless time \maths{\mu\,T/\hbar} elapsed after the occurrence of the quench, as given in Eqs.~\eqref{Eq:AutocorrelationFunctionQuinquies} and \eqref{Eq:LongRangePlateau} supplemented by Eqs.~\eqref{Eq:DensityCorrelationsResult}, \eqref{Eq:SoundSpeedRelativeShiftResult}, \eqref{Eq:LightConeBoundary}, and \eqref{Eq:EffectiveTemperature} for \maths{\mathcal{V}/\mu=0.5} and \maths{\sigma/\xi=1}; the abscissas of the curves' siding edges equal \maths{\zeta_{T}/\xi=(2\,\mu\,T/\hbar)\,(1+\Delta\tilde{s}/s)} [cf.~Eqs.~\eqref{Eq:LightConeBoundary} and use \maths{s/\xi=\mu/\hbar}]. Dashed curves: Corresponding behaviors in the strict absence of disorder, that is, when \maths{\mathcal{V}=0}; in this case, \maths{\Delta\tilde{s}=0} and the abscissas of the curves' siding edges equal \maths{2\,\mu\,T/\hbar}.}
\label{Fig:AutocorrelationFunction}
\end{figure}

In Fig.~\ref{Fig:AutocorrelationFunction}, we plot \maths{(c_{0}/v_{1})\,\rho_{\mathrm{in}}\,\xi\;\ln[g^{(1)}(\zeta-\zeta')/\rho_{\mathrm{in}}]} as a function of \maths{|\zeta-\zeta'|/\xi} for different values of (i) \maths{\mu\,T/\hbar\gg1} but fixed values of (ii) \maths{\mathcal{V}/\mu\ll1} and (iii) \maths{\sigma/\xi\ll\mu\,T/\hbar}. The condition (i) is the limit of long postquench duration discussed in the last paragraph of Sec.~\ref{SubSec:InputOutputRelations}. The condition (ii) is the limit of weak disorder assumed from the third paragraph of Sec.~\ref{SubSec:GrossPitaevskiiClassicalField}. In the limit (iii) finally, the system feels the presence of a sufficient number of random scatterers after the occurrence of the quench so to consider the effect of the disordered potential relevant.

At \maths{\tau=0^{+}}, right after the quench, one may show that the coherence function of \maths{\hat{\Psi}(\zeta,\tau)} equals \maths{(\rho_{\mathrm{in}})^{-}} for all \maths{|\zeta-\zeta'|}. This means that the beam of light \maths{\boldsymbol{1}} remains as fully coherent as before entering the fiber. The \maths{g^{(1)}} function starts being affected by the quench a significant duration \maths{T} after its occurrence. Focusing on one of the solid curves of Fig.~\ref{Fig:AutocorrelationFunction}, three regimes depending on \maths{|\zeta-\zeta'|} may be identified. At very short ranges, \maths{g^{(1)}} displays a nontrivial \maths{|\zeta-\zeta'|} dependence given in Eq.~\eqref{Eq:AutocorrelationFunctionVeryShortRanges}. Afterwards and up to \maths{|\zeta-\zeta'|=\zeta_{T}}, its natural logarithm linearly decays, which corresponds to an exponential decay for \maths{g^{(1)}}. This interesting regime is entirely described by the first row in the right-hand side of Eq.~\eqref{Eq:AutocorrelationFunctionQuinquies} and is discussed in detail below. For \maths{|\zeta-\zeta'|>\zeta_{T}} finally, the \maths{g^{(1)}} function no longer depends on \maths{|\zeta-\zeta'|}. Its constant value is given in Eq.~\eqref{Eq:LongRangePlateau} and is also subjected to a discussion in the next paragraphs.

As \maths{T} increases, \maths{\zeta_{T}} is pushed to larger values of \maths{|\zeta-\zeta'|} and the long-range, \maths{|\zeta-\zeta'|>\zeta_{T}}, plateau of the \maths{g^{(1)}} function decreases, which we will discuss later. This evolution continues until the system reaches, in the limit \maths{\mu\,T/\hbar=\infty}, a state where \maths{g^{(1)}(\zeta-\zeta')} is exponential across the whole 1D system [putting aside its short-range behavior \eqref{Eq:AutocorrelationFunctionVeryShortRanges}]:
\begin{subequations}
\label{Eq:PrethermalizedState}
\begin{align}
\label{Eq:PrethermalizedState-a}
g^{(1)}(\zeta-\zeta')&\propto\rho_{\mathrm{in}}\exp\!\bigg[{-}\pi\,\frac{v_{1}}{c_{0}}\,\frac{|\zeta-\zeta'|}{\rho_{\mathrm{in}}\,\Lambda^{2}(\tilde{\theta}_{\mathrm{eff}})}\bigg], \\
\label{Eq:PrethermalizedState-b}
\Lambda(\tilde{\theta}_{\mathrm{eff}})&=\bigg(\frac{2\pi\,\hbar^{2}}{m\,\tilde{\theta}_{\mathrm{eff}}}\bigg)^{\frac{1}{2}}.
\end{align}
\end{subequations}
Apart from the velocity ratio \maths{v_{1}/c_{0}}, Eq.~\eqref{Eq:PrethermalizedState-a} is strictly identical to the long-range \maths{g^{(1)}} function of a dilute gas of thermal, weakly interacting boson atoms in the 1D degenerate regime \maths{1/\rho_{\mathrm{in}}\ll\Lambda(\tilde{\theta}_{\mathrm{eff}})}, with \maths{\rho_{\mathrm{in}}} the uniform density of the gas and \maths{\Lambda(\tilde{\theta}_{\mathrm{eff}})} the thermal de Broglie wavelength \cite{Petrov2003, NoteIdealGas}. As a result, the state described above may be interpreted as the thermal equilibrium state of the system, reached a long time after the quench.

\begin{figure}[t!]
\includegraphics[width=\linewidth]{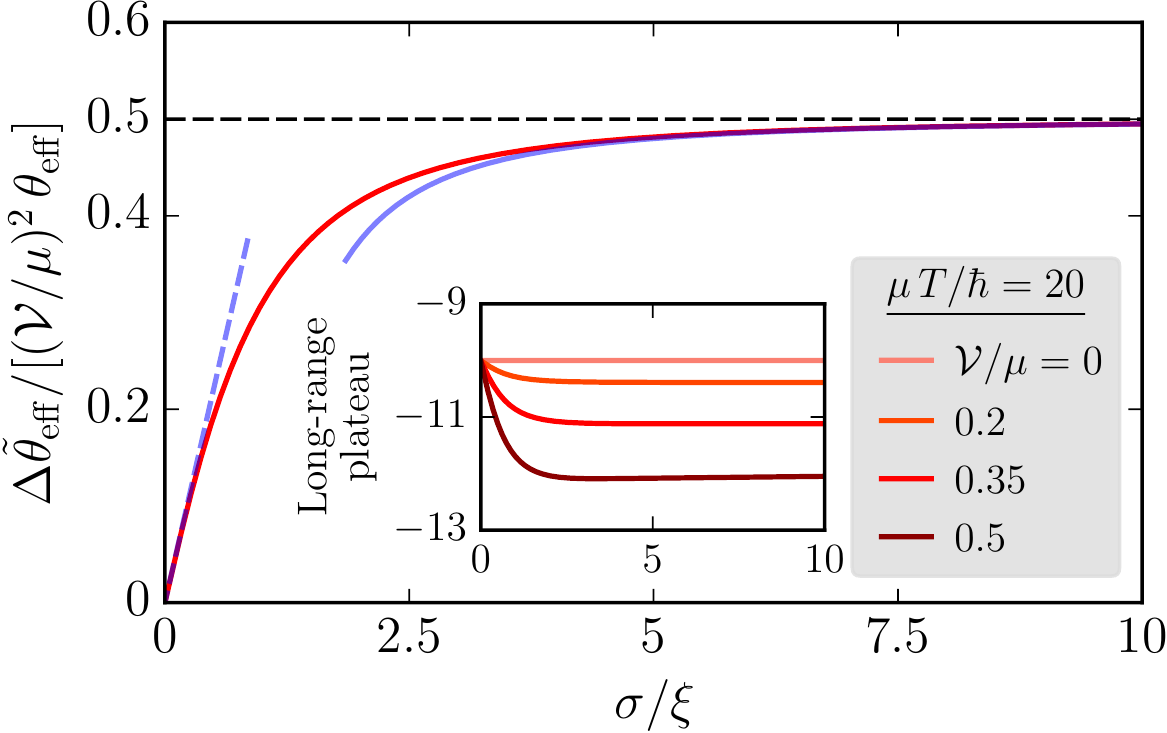}
\caption{(Color online) Main plot, red curve: Normalized disorder-induced correction to the prethermalization temperature as a function of \maths{\sigma/\xi}, as given in Eqs.~\eqref{Eq:EffectiveTemperature-b}. Main plot, blue curves: Asympotic behaviors when \maths{\sigma/\xi\ll1} (dashed curve) and when \maths{\sigma/\xi\gg1} (solid curve), as given in Eq.~\eqref{Eq:EffectiveTemperatureAsymptoticResult}. Inset: Corresponding long-range, \maths{|\zeta-\zeta'|>\zeta_{T}}, plateau of \maths{(c_{0}/v_{1})\,\rho_{\mathrm{in}}\,\xi\;\ln[g^{(1)}(\zeta-\zeta')/\rho_{\mathrm{in}}]} as a function of \maths{\sigma/\xi} for different values of \maths{\mathcal{V}/\mu}, as given in Eq.~\eqref{Eq:LongRangePlateau} supplemented by Eqs.~\eqref{Eq:DensityCorrelationsResult}, \eqref{Eq:SoundSpeedRelativeShiftResult}, and \eqref{Eq:LightConeBoundary} for \maths{\mu\,T/\hbar=20}.}
\label{Fig:TemperaturePlateau}
\end{figure}

The effective temperature \maths{\tilde{\theta}_{\mathrm{eff}}} of this thermalized state is explicitly formulated in Eqs.~\eqref{Eq:EffectiveTemperature}. In the absence of disorder, \maths{\tilde{\theta}_{\mathrm{eff}}=\theta_{\mathrm{eff}}=\mu/2=g\,\bar{\rho}_{0}/2} is nothing but the mean interaction energy of the photons in the fiber, which is the usual energy deposed by a steplike interaction quench in a clean quantum nonlinear Schr\"odinger system (see, e.g., Refs.~\cite{Kitagawa2011, Gring2012, Kuhnert2013}). In the presence of disorder, it is natural that the energy provided by the quench, and so \maths{\tilde{\theta}_{\mathrm{eff}}}, are enhanced with respect to the configuration without disorder since the quench also involves the disordered-potential term in Eq.~\eqref{Eq:QuantumNLSFormalismBis-a}. At the second order in the weak-disorder parameter \maths{\mathcal{V}/\mu\ll1}, the disorder-modified \maths{\tilde{\theta}_{\mathrm{eff}}} turns out to be positively shifted from its unperturbed counterpart \maths{\theta_{\mathrm{eff}}} by the random fluctuations \maths{G(0)=\overline{\delta\rho_{0}^{2}(\zeta)}} of the mean density of photons in the fiber, as shown by the first of the equalities \eqref{Eq:EffectiveTemperature-b}. The second one provides its explicit \maths{\sigma/\xi} dependence. In Fig.~\ref{Fig:TemperaturePlateau}, we plot \maths{\Delta\tilde{\theta}_{\mathrm{eff}}/[(\mathcal{V}/\mu)^{2}\,\theta_{\mathrm{eff}}]} as a function of \maths{\sigma/\xi} and indicate its asymptotic \maths{\sigma/\xi\ll1} and \maths{\sigma/\xi\gg1} behaviors given in Eq.~\eqref{Eq:EffectiveTemperatureAsymptoticResult}. The fact that \maths{\Delta\tilde{\theta}_{\mathrm{eff}}\propto\mathcal{V}^{2}\,\sigma} when \maths{\sigma/\xi\ll1} and that \maths{\Delta\tilde{\theta}_{\mathrm{eff}}\propto\mathcal{V}^{2}} when \maths{\sigma/\xi=\infty} can be understood in the same way as for \maths{\Delta\tilde{s}} (see Sec.~\ref{SubSec:BogoliubovQuantumFluctuations}).

Importantly, the present Bogoliubov approach of quantum fluctuations accounts neither for the interactions between the excitations of the quantum fluid of light nor for the interactions between these excitations and the classical background fluid. Such interactions are nevertheless expected to occur at very long times, leading to damping in the many-body quantum system. In this respect, although referred to as thermalized, the thermal state described in the two previous paragraphs does not correspond to the actual thermal equilibrium state of the postquench system but rather to some quasistationary intermediate thermal state usually referred to as prethermalized \cite{Berges2004, Marino2012, VanDenWorm2013, Marcuzzi2013, Marino2014, Moeckel2008, Eckstein2009, Moeckel2009, Moeckel2010, Kitagawa2011, Barnett2011, Gring2012, Kuhnert2013, Langen2013, Mitra2013, Buchhold2016}. The investigation of true thermalization (although possibly blocked by localization phenomena in the presence of disorder \cite{Gornyi2005, Basko2006, Nandkishore2015, Altman2015}) requires to go beyond Bogoliubov's theory, at best to account for the full many-body quantum dynamics \eqref{Eq:QuantumNLSFormalismBis}. This goes beyond the scope of the present manuscript but will be subjected to future works.

According to what precedes, a given point \maths{\zeta} in the disordered 1D quantum fluid of light establishes thermal correlations with another point \maths{\zeta'} as long as their separation distance \maths{|\zeta-\zeta'|} is smaller than the characteristic length \maths{\zeta_{T}} given in Eqs.~\eqref{Eq:LightConeBoundary}. The latter linearly scales with the time \maths{T} elapsed after the quench as well as with the disorder-renormalized Bogoliubov speed of sound \maths{\tilde{s}} given in Eqs.~\eqref{Eq:PhononDispersionRelation}. This means that the prethermalized state emerges in a light-cone way in the system, \maths{\tilde{s}} being the characteristic propagation velocity of the thermal correlations. This may be understood as follows. From the entrance of the nonlinear optical fiber, a spontaneous degenerate four-wave mixing occurs in the beam of light of carrier angular frequency \maths{\omega_{1}}, with two coherently correlated sidebands symmetrically peaked around \maths{\omega_{1}+\Delta_{\mathrm{s}}>\omega_{1}} (the signal) and \maths{\omega_{1}+\Delta_{\mathrm{i}}=\omega_{1}-\Delta_{\mathrm{s}}<\omega_{1}} (the idler) \cite{Agrawal2013}. Within the \maths{z\longleftrightarrow t} mapping, this equivalently means that the quench-induced excitation process of the 1D quantum fluid of light consists in the spontaneous emission of coherently correlated Bogoliubov fluctuations with opposite wavenumbers \maths{k_{\mathrm{s}}=-\Delta_{\mathrm{s}}/v_{1}=-k<0} and \maths{k_{\mathrm{i}}=-\Delta_{\mathrm{i}}/v_{1}=\Delta_{\mathrm{s}}/v_{1}=k>0} along the \maths{\zeta=v_{1}\,t-z} axis [see the definition of \maths{k} as a function of \maths{\Delta} right before Eq.~\eqref{Eq:MatterFieldFreeSpaceBis-a}]. These quasiparticles propagate faster than the sound waves, even in the presence of disorder (provided the latter is not too strong). As a result, if a Bogoliubov excitation with \maths{k} and another one with \maths{-k} are respectively located at \maths{\zeta} and \maths{\zeta'} a time \maths{T} after the quench, the two must be separated at least by
\begin{equation}
\label{Eq:LightConeBoundaryPhonons}
|\zeta-\zeta'|_{\mathrm{sound}}=|\tilde{s}\,T-(-\tilde{s}\,T)|=2\,\tilde{s}\,T=\zeta_{T}
\end{equation}
to be coherently correlated. As a consequence, coherence cannot exist for \maths{|\zeta-\zeta'|<\zeta_{T}}. In this case, it is automatically replaced with thermal correlations since the quench effectively heats the system. Such a light-cone-like correlation spreading is a widely observed phenomenon, e.g., in cold atomic vapors \cite{Cheneau2012, Langen2013}, condensed-matter \cite{Lieb1972, Cramer2008, Lauchli2008, Mathey2010, Carleo2014, Geiger2014, Cevolani2016, Cevolani2017}, quantum field \cite{Calabrese2006, Calabrese2007, Chiocchetta2016-b}, and quantum information \cite{Bravyi2006} theory.

As mentioned before, the \maths{g^{(1)}} function no longer decays when \maths{|\zeta-\zeta'|>\zeta_{T}}. In this case, it is locked to the value it takes at \maths{|\zeta-\zeta'|=\zeta_{T}}, given in Eq.~\eqref{Eq:LongRangePlateau} in logarithmic units. At a finite \maths{T}, even though this long-range plateau is nonzero, it is nevertheless smaller than \maths{\rho_{\mathrm{in}}}. This indicates that the system partially lost its coherence after the occurrence of the quench, precisely due to the concomitant generation of thermal fluctuations, as detailed in the last paragraphs. As \maths{T} increases, it is pushed to zero until the system becomes fully incoherent in the limiting case where \maths{\mu\,T/\hbar=\infty}. In Fig.~\ref{Fig:TemperaturePlateau}, we fix the value of \maths{\mu\,T/\hbar} and plot the corresponding long-range plateau of \maths{(c_{0}/v_{1})\,\rho_{\mathrm{in}}\,\xi\;\ln[g^{(1)}(\zeta-\zeta')/\rho_{\mathrm{in}}]} as a function of \maths{\sigma/\xi} for different values of \maths{\mathcal{V}/\mu}. The loss of long-range coherence is as significant as the amplitude of the disordered potential is large, as intuition suggests.

\section{On the quantum nature of the decoherence and orders of magnitude}
\label{Sec:OnTheQuantumNatureOfTheDecoherenceAndOrdersOfMagnitude}

The quench-induced loss of macroscopic coherence we predict is a quantum effect that stems from the mismatch of the vacua of the quantum light field between the exterior of the fiber (free-space vacuum) and the interior (Bogoliubov vacuum). This can be  explicitly seen, already in the absence of disorder (\maths{\mathcal{V}=0}), by reformulating Eq.~\eqref{Eq:LongRangePlateau} in terms of the original optical parameters of the problem. This gives
\begin{subequations}
\label{Eq:LongRangePlateauNumerics}
\begin{align}
\label{Eq:LongRangePlateauNumerics-a}
&\left.\frac{g^{(1)}(|\zeta-\zeta'|>\zeta_{T})}{\rho_{\mathrm{in}}}=e^{-L/L_{\mathrm{c}}},\right. \\
\label{Eq:LongRangePlateauNumerics-b}
&\left.L_{\mathrm{c}}=\frac{1}{(8\,\pi^{5})^{\frac{1}{2}}\,\hbar\,c_{0}}\,\bigg[\frac{\lambda_{1}^{5}\,\mathcal{A}_{1}^{3}\,|D_{1}^{\vphantom{3}}|}{(P_{1})_{\mathrm{in}}\,|(\tilde{n}_{2})_{1}|^{3}}\bigg]^{\frac{1}{2}},\right.
\end{align}
\end{subequations}
where \maths{\lambda_{1}=2\pi\,c_{0}/\omega_{1}} is the wavelength of the beam \maths{\boldsymbol{1}} in free space, \maths{(P_{1})_{\mathrm{in}}=\frac{1}{2}\,c_{0}\,\varepsilon_{0}\,|A_{1}(0^{-},t)|^{2}} is its input power, and \maths{(\tilde{n}_{2})_{1}=(n_{2})_{1}/[\frac{1}{2}\,c_{0}\,\varepsilon_{0}\,(n_{\mathrm{L}})_{1}]} is the Kerr-nonlinearity coefficient at \maths{\omega_{1}} in intensity units. The proportionality of the inverse of the coherence length \maths{L_\mathrm{c}} to \maths{\hbar} signals the quantum nature of the decoherence mechanism described in this paper [technically, this \maths{\hbar} dependence stems from the finite value of the quantum commutator \eqref{Eq:QuantumFormalism-b}]. In the classical limit \maths{\hbar\to0}, \maths{e^{-L/L_{\mathrm{c}}}\to1}, so that the light field remains fully coherent upon crossing the fiber.

At weak disorder, Eqs.~\eqref{Eq:LongRangePlateauNumerics} also provide a good numerical estimate for the quench-induced loss of macroscopic coherence. As an example, we consider a \maths{\mathcal{A}_{1}=10\;\mu\mathrm{m}^{2}}-thick, \maths{L=500\;\mathrm{km}}-long silica-based telecom fiber illuminated by an infrared laser beam of wavelength \maths{\lambda_{1}=1.55\;\mu\mathrm{m}} and peak-power \maths{(P_{1})_{\mathrm{in}}=1\;\mathrm{kW}}. In this case, the group-velocity-dispersion parameter \maths{|D_{1}|\simeq27.95\;\mathrm{ps}^{2}\cdot\mathrm{km}^{-1}} and the Kerr-nonlinearity coefficient \maths{|(\tilde{n}_{2})_{1}|\simeq3.46\times10^{-5}\;\mu\mathrm{m}^{2}\cdot\mathrm{kW}^{-1}} \cite{Agrawal2013} [which gives \maths{|(\tilde{n}_{2})_{1}|\,(P_{1})_{\mathrm{in}}/\mathcal{A}_{1}\simeq3.46\times10^{-6}} for the nonlinear shift of the refractive index]. We then find a decoherence \maths{1-g^{(1)}(|\zeta-\zeta'|>\zeta_{T})/\rho_{\mathrm{in}}\simeq1\%}. It is as expected small, since of quantum origin, but it may be enhanced to almost 30\% by taking a \maths{(\tilde{n}_{2})_{1}} barely 10 times larger than the standard Kerr-nonlinearity coefficient for silica. Such an order of magnitude for \maths{(\tilde{n}_{2})_{1}} is relatively commonly encountered, e.g., in silicon photonics \cite{Pavesi2004}.

Note that the approach presented in this paper could be used as well to describe a classical nonequilibrium dynamics through a nonlinear fiber. In this case, the fluctuations around the monochromatic plane-wave pump in Eq.~\eqref{Eq:ElectricFieldFreeSpace} would be of purely classical origin. Such fluctuations are naturally present in most experiments and should give rise to a loss of coherence as well, but of classical origin. Due to the \maths{\hbar}-independence of these classical fluctuations, we suspect the effect to be observable on propagation distances shorter than the length scales needed to see quantum effects at standard optical nonlinearity (compare, e.g., Refs.~\cite{Chiocchetta2016-a} and \cite{Connaughton2005, Sun2012, Santic2018} on the kinetic relaxation of a quantum or classical beam of nonlinear light).

\vfill

\section{Conclusion}
\label{Sec:Conclusion}

Using a general theory for the quantum propagation of a paraxial beam of quasimonochromatic light in a dispersive, inhomogeneous, and nonlinear dielectric medium, we investigated the quantum coherence of a randomly cross-phase modulated quantum light field emerging from a lossless 1D optical fiber with a quadratic dispersion relation and a local Kerr nonlinearity.

In this theory, the space propagation of the quantum field is mapped onto a time evolution and the actual time parameter is identified as a space coordinate. The group-velocity dispersion of the fiber generates a mass term. The instantaneous power of the auxiliary beam of light responsible for cross-phase modulation serves as a spatially correlated disordered potential once properly randomized as a function of time by means of a light modulator. Finally, self-phase modulation provides effective photon-photon interactions.

In this all-optical setup, we entirely reformulated our predictions in the language of many-body quantum physics, precisely in terms of the response of a 1D quantum fluid of light to a disorder and interaction quench at the entrance of the fiber. Our work then illustrates the interest of nonlinear photonics as a powerful platform for simulating the postquench nonequilibrium dynamics of disordered many-body quantum systems.

We specifically focused on the case where both the disordered potential and the photon-photon interactions are weak. At the exit of the fiber, the coherence function of the transmitted quantum field features a peculiar prethermalization dynamics in the disordered 1D quantum fluid of light. As a result of the quench, thermal correlations are produced in the system, spreading in a light-cone way at the disorder-renormalized Bogoliubov speed of sound. They are exponentially distributed with a temperature depending on the disorder and the interaction energies deposed by the quench, and their emergence is accompanied with a disorder-dependent loss of macroscopic coherence. This highlights a fundamental limit to the coherent propagation of a quantum-fluctuating light along a 1D nonlinear fiber.

\begin{acknowledgments}
We acknowledge Matthieu Bellec and Tony Prat for interesting discussions. This work was successively supported by the Centre National de la Recherche Scientifique (CNRS), by the Agence Nationale de la Recherche (ANR) under the grant ANR-14-CE26-0032 LOVE, and by the Universit\'e de Cergy-Pontoise.
\end{acknowledgments}

\appendix

\section{Cross-phase modulation from a single-beam effective model}
\label{App:CrossPhaseModulationFromASingleBeamEffectiveModel}

Equation \eqref{Eq:PropagationEquation} may be derived within the framework of a specific single-beam phenomenological model. Within this model, the cross-phase modulation induced by the beam of light \maths{\boldsymbol{3-\alpha}=\boldsymbol{2}~\text{or}~\boldsymbol{1}} is encapsulated in the linear refractive index seen by the beam of light \maths{\boldsymbol{\alpha}=\boldsymbol{1}~\text{or}~\boldsymbol{2}}, so that the latter perceives the overall effective refractive index
\begin{align}
\notag
n_{\mathrm{eff}}(\mathbf{r},\omega)&\left.=n_{\mathrm{L}}(\omega)+\Delta n_{\mathrm{L}}(x,y,\omega)+(\Delta n_{\mathrm{L}})'(\mathbf{r},\omega)\right. \\
\label{Eq:EffectiveDielectric}
&\left.\hphantom{=}+\Delta n_{\mathrm{NL}}(\mathbf{r},\omega).\right.
\end{align}
In Eq.~\eqref{Eq:EffectiveDielectric}, the subscripts ``\maths{\mathrm{L}}'' and ``\maths{\mathrm{NL}}'' refer to the contributions to the envelope \maths{\mathcal{D}_{\alpha}(\mathbf{r},t)=\varepsilon_{0}\,[n_{\mathrm{eff}}(\mathbf{r},\omega_{\alpha})]^{2}\,\mathcal{E}_{\alpha}(\mathbf{r},t)} of \maths{\boldsymbol{\alpha}}'s complex electric displacement field that are respectively ``linear'' and ``nonlinear'' in \maths{\mathcal{E}_{\alpha}(\mathbf{r},t)}. The \maths{\omega}-dependent refractive index \maths{n_{\mathrm{L}}(\omega)=(\omega/c_{0})^{-1}\,k(\omega)} is responsible for the chromatic dispersion of the fiber's core. Its \maths{x}- and \maths{y}-dependent correction \maths{\Delta n_{\mathrm{L}}(x,y,\omega)} originates from the refractive-index mismatch between the core and the cladding and makes \maths{\boldsymbol{\alpha}} transversally confined, aligned with the \maths{z} axis. The second linear shift---which phenomenologically describes cross-phase modulation---is explicitly defined as
\begin{equation}
\label{Eq:XPMRefractiveIndex}
(\Delta n_{\mathrm{L}})'(\mathbf{r},\omega)=2\,n_{2}(\omega)\,|\mathcal{E}_{3-\alpha}(\mathbf{r},t)|^{2}.
\end{equation}
Finally, \maths{\Delta n_{\mathrm{NL}}(\mathbf{r},\omega)=n_{2}(\omega)\,|\mathcal{E}_{\alpha}(\mathbf{r},t)|^{2}} is the usual nonlinear contribution responsible for self-phase modulation.

Writing Maxwell's equations for the single beam of light \maths{\boldsymbol{\alpha}} and assuming that \maths{\Delta n_{\mathrm{L}}(x,y,\omega)}, \maths{(\Delta n_{\mathrm{L}})'(\mathbf{r},\omega)}, and \maths{\Delta n_{\mathrm{NL}}(\mathbf{r},\omega)} are small compared to \maths{n_{\mathrm{L}}(\omega)} in the expansion \eqref{Eq:EffectiveDielectric}, it is straightforward to show that the time Fourier transform
\begin{subequations}
\label{Eq:ElectricFieldFrequencyDomain}
\begin{align}
\label{Eq:ElectricFieldFrequencyDomain-a}
E_{\alpha}(\mathbf{r},\omega)&=\int dt\,E_{\alpha}(\mathbf{r},t)\,e^{i\omega t} \\
\label{Eq:ElectricFieldFrequencyDomain-b}
&=F_{\alpha}(x,y)\,A_{\alpha}(z,\omega-\omega_{\alpha})\,e^{ik_{\alpha}z}
\end{align}
\end{subequations}
of \maths{\boldsymbol{\alpha}}'s complex electric field \eqref{Eq:ElectricField} satisfies the following Helmholtz equation:
\begin{equation}
\label{Eq:HelmholtzEquation}
{\frac{\partial^{2}E_{\alpha}}{\partial x^{2}}+\frac{\partial^{2}E_{\alpha}}{\partial y^{2}}+\frac{\partial^{2}E_{\alpha}}{\partial z^{2}}+[n_{\mathrm{eff}}(\mathbf{r},\omega)\,\omega/c_{0}]^{2}\,E_{\alpha}=0.}
\end{equation}
Plugging Eq.~\eqref{Eq:ElectricFieldFrequencyDomain-b} into Eq.~\eqref{Eq:HelmholtzEquation}, one gets
\begin{align}
\notag
&\left.\bigg(\frac{\partial^{2}F_{\alpha}}{\partial x^{2}}+\frac{\partial^{2}F_{\alpha}}{\partial y^{2}}\bigg)\,A_{\alpha}+F_{\alpha}\,\bigg(\frac{\partial^{2}A_{\alpha}}{\partial z^{2}}+2\,i\,k_{\alpha}\,\frac{\partial A_{\alpha}}{\partial z}\bigg)\right. \\
\label{Eq:HelmholtzEquationBis}
&\left.{\quad}+F_{\alpha}\,\{[n_{\mathrm{eff}}(\mathbf{r},\omega)\,\omega/c_{0}]^{2}-(k_{\alpha})^{2}\}\,A_{\alpha}=0,\right.
\end{align}
where \maths{A_{\alpha}=A_{\alpha}(z,\omega-\omega_{\alpha})}.

Making use of Eq.~\eqref{Eq:EffectiveDielectric} as a perturbative expansion around \maths{n_{\mathrm{L}}(\omega)}, of \maths{k(\omega)=n_{\mathrm{L}}(\omega)\,\omega/c_{0}} given in Eq.~\eqref{Eq:DispersionRelation}, and remembering that \maths{A_{\alpha}(z,\omega-\omega_{\alpha})} slowly varies over scales \maths{\sim2\pi/k_{\alpha}} and is strongly peaked around \maths{\omega_{\alpha}}, we reduce Eq.~\eqref{Eq:HelmholtzEquationBis} to, after projection onto \maths{F_{\alpha}(x,y)},
\begin{align}
\notag
i\,\frac{\partial A_{\alpha}}{\partial z}&\left.=-\frac{D_{\alpha}}{2}\,(\omega-\omega_{\alpha})^{2}\,A_{\alpha}-\frac{1}{v_{\alpha}}\,(\omega-\omega_{\alpha})\,A_{\alpha}\right. \\
\notag
&\left.\hphantom{=}-\frac{\omega_{\alpha}}{c_{0}}\int dx\,dy\;\Delta n_{\mathrm{NL}}(\mathbf{r},\omega_{\alpha})\,|F_{\alpha}|^{2}\,A_{\alpha}\right. \\
\notag
&\left.\hphantom{=}-\frac{\omega_{\alpha}}{c_{0}}\int dx\,dy\;(\Delta n_{\mathrm{L}})'(\mathbf{r},\omega_{\alpha})\,|F_{\alpha}|^{2}\,A_{\alpha}\right. \\
\notag
&\left.\hphantom{=}+\int dx\,dy\,\bigg[\frac{1}{2\,k_{\alpha}}\,\bigg(\bigg|\frac{\partial F_{\alpha}}{\partial x}\bigg|^{2}+\bigg|\frac{\partial F_{\alpha}}{\partial y}\bigg|^{2}\bigg)\right. \\
\label{Eq:HelmholtzEquationTer}
&\left.\hphantom{=}-\frac{\omega_{\alpha}}{c_{0}}\,\Delta n_{\mathrm{L}}(x,y,\omega_{\alpha})\,|F_{\alpha}|^{2}\bigg]\,A_{\alpha}.\right.
\end{align}
Using \maths{\Delta n_{\mathrm{NL}}(\mathbf{r},\omega_{\alpha})=(n_{2})_{\alpha}\,|F_{\alpha}(x,y)|^{2}\,|A_{\alpha}(z,t)|^{2}}, \maths{(\Delta n_{\mathrm{L}})'(\mathbf{r},\omega_{\alpha})=2\,(n_{2})_{\alpha}\,|F_{3-\alpha}(x,y)|^{2}\,|A_{3-\alpha}(z,t)|^{2}}, and absorbing the last double integral, \maths{A_{\alpha}} independent, into the phase of \maths{A_{\alpha}(z,\omega-\omega_{\alpha})}, we eventually obtain Eq.~\eqref{Eq:PropagationEquation} in the time Fourier domain.

Note that \maths{\Delta n_{\mathrm{NL}}(\mathbf{r},\omega_{\alpha})} and \maths{(\Delta n_{\mathrm{L}})'(\mathbf{r},\omega_{\alpha})} defined above are time dependent, which may a priori appear suspect within the present angular-frequency derivation. This time dependence can be taken into account as we presently do if the spectral bandwidths of the pulses \maths{\boldsymbol{\alpha}} and \maths{\boldsymbol{3-\alpha}} are small fractions of \maths{\omega_{\alpha}} and \maths{\omega_{3-\alpha}} \cite{Agrawal2013}, which is actually the case here.

\section{Flux of the Poynting vector of a cross-phase modulated optical beam}
\label{App:FluxOfThePoyntingVectorOfACrossPhaseModulatedOpticalBeam}

In this appendix, we derive an expression for the flux of \maths{\boldsymbol{1}}'s Poynting vector within the single-beam effective model used in the text and detailed in Appendix \ref{App:CrossPhaseModulationFromASingleBeamEffectiveModel}.

Taking the electric field \maths{\mathbf{E}_{1}(\mathbf{r},t)} along the \maths{x} axis and the magnetic field \maths{\mathbf{H}_{1}(\mathbf{r},t)} along the \maths{y} axis, i.e.,
\begin{align}
\label{Eq:RealElectricField}
\mathbf{E}_{1}(\mathbf{r},t)&=\mathrm{Re}[\mathcal{E}_{1}(\mathbf{r},t)\,e^{i(k_{1}z-\omega_{1}t)}]\,\hat{\mathbf{x}}, \\
\label{Eq:RealMagneticField}
\mathbf{H}_{1}(\mathbf{r},t)&=\mathrm{Re}[\mathcal{H}_{1}(\mathbf{r},t)\,e^{i(k_{1}z-\omega_{1}t)}]\,\hat{\mathbf{y}},
\end{align}
the Poynting vector \maths{\mathbf{\Pi}_{1}(\mathbf{r},t)=\mathbf{E}_{1}(\mathbf{r},t)\times\mathbf{H}_{1}(\mathbf{r},t)} gets aligned along the \maths{z} axis:
\begin{subequations}
\label{Eq:PoyntingVector}
\begin{align}
\label{Eq:PoyntingVector-a}
\mathbf{\Pi}_{1}(\mathbf{r},t)&\left.=\mathrm{Re}[\Pi_{1}(\mathbf{r},t)]\,\hat{\mathbf{z}},\right. \\
\notag
\Pi_{1}(\mathbf{r},t)&\left.=\frac{1}{2}\,\mathcal{E}_{1}^{\ast}(\mathbf{r},t)\,[\mathcal{H}_{1}^{\vphantom{\ast}}(\mathbf{r},t)\right. \\
\label{Eq:PoyntingVector-b}
&\left.\hphantom{=}+\mathcal{H}_{1}^{\ast}(\mathbf{r},t)\,e^{-2i(k_{1}z-\omega_{1}t)}].\right.
\end{align}
\end{subequations}

In the experiment, the photodetectors are not able to instantaneously record \eqref{Eq:PoyntingVector} but instead perform an average over a few, at least one, time period(s) \maths{2\pi/\omega_{1}} of the carrier wave. As a result, we now make use of the corresponding time-average angle brackets \maths{\langle{\cdots}\rangle_{t}=(2\pi/\omega_{1})^{-1}\int_{0}^{2\pi/\omega_{1}}dt\,({\cdots})}. Since the envelopes \maths{\mathcal{E}_{1}(\mathbf{r},t)} and \maths{\mathcal{H}_{1}(\mathbf{r},t)} are almost \maths{t} independent over a duration of the order of \maths{2\pi/\omega_{1}}, we readily get
\begin{align}
\label{Eq:PoyntingVectorBis}
\langle\Pi_{1}(\mathbf{r},t)\rangle_{t}\simeq\frac{1}{2}\,\mathcal{E}_{1}^{\ast}(\mathbf{r},t)\,\mathcal{H}_{1}^{\vphantom{\ast}}(\mathbf{r},t).
\end{align}
The \maths{\mathcal{E}_{1}(\mathbf{r},t)} dependence of \maths{\mathcal{H}_{1}(\mathbf{r},t)} may be obtained from Maxwell-Amp\`ere's law in the slowly-varying-envelope approximation, i.e.,
\begin{equation}
\label{Eq:MaxwellAmpereLaw}
\mathcal{H}_{1}(\mathbf{r},t)\simeq\frac{\omega_{1}}{k_{1}}\,\mathcal{D}_{1}(\mathbf{r},t)=\frac{c_{0}}{(n_{\mathrm{L}})_{1}}\,\mathcal{D}_{1}(\mathbf{r},t).
\end{equation}
In this equation,
\begin{subequations}
\label{Eq:ElectricDisplacementField}
\begin{align}
\notag
\mathcal{D}_{1}(\mathbf{r},t)&\left.=\varepsilon_{0}\int\frac{d\omega}{2\pi}\,[n_{\mathrm{eff}}(\mathbf{r},\omega)]^{2}\right. \\
\label{Eq:ElectricDisplacementField-a}
&\left.\hphantom{=}\times\mathcal{E}_{1}(\mathbf{r},\omega-\omega_{1})\,e^{-i(\omega-\omega_{1})t}\right. \\
\label{Eq:ElectricDisplacementField-b}
&\left.\simeq\varepsilon_{0}\,[n_{\mathrm{eff}}(\mathbf{r},\omega_{1})]^{2}\,\mathcal{E}_{1}(\mathbf{r},t)\right.
\end{align}
\end{subequations}
is the envelope of the complex electric displacement field. Inserting \eqref{Eq:ElectricDisplacementField-b} into \eqref{Eq:MaxwellAmpereLaw} and then \eqref{Eq:MaxwellAmpereLaw} into \eqref{Eq:PoyntingVectorBis}, we end up with
\begin{equation}
\label{Eq:PoyntingVectorTer}
\langle\Pi_{1}(\mathbf{r},t)\rangle_{t}\simeq\frac{1}{2}\,c_{0}\,\varepsilon_{0}\,\frac{[n_{\mathrm{eff}}(\mathbf{r},\omega_{1})]^{2}}{(n_{\mathrm{L}})_{1}}\,|\mathcal{E}_{1}(\mathbf{r},t)|^{2}.
\end{equation}
From now on, we implicitly neglects the \maths{\Delta n_{\mathrm{NL}}(\mathbf{r},\omega_{1})\propto|\mathcal{E}_{1}(\mathbf{r},t)|^{2}} dependence of \maths{n_{\mathrm{eff}}(\mathbf{r},\omega_{1})} so that the right-hand side of Eq.~\eqref{Eq:PoyntingVectorTer} is simply proportional to \maths{|\mathcal{E}_{1}(\mathbf{r},t)|^{2}}. This is a good approximation provided both the Kerr-nonlinearity coefficient and \maths{\boldsymbol{1}}'s intensity are small. With this, the flux of \eqref{Eq:PoyntingVectorTer} is expressed as
\begin{align}
\notag
&\left.\int dx\,dy\,\langle\Pi_{1}(\mathbf{r},t)\rangle_{t}\right. \\
\label{Eq:FluxPoyntingVectorFiberClassical}
&\left.{\quad}\simeq\frac{1}{2}\,c_{0}\,\varepsilon_{0}\,(n_{\mathrm{L}})_{1}\,\mathcal{F}(v_{1}\,t-z)\,|A_{1}(z,t)|^{2},\right.
\end{align}
where
\begin{align}
\notag
\mathcal{F}(v_{1}\,t-z)&\left.=\int dx\,dy\,\bigg[1+\frac{\Delta n_{\mathrm{L}}(x,y,\omega_{1})}{(n_{\mathrm{L}})_{1}}\right. \\
\label{Eq:StructureFactor}
&\left.\hphantom{=}+\frac{(\Delta n_{\mathrm{L}})'(\mathbf{r},\omega_{1})}{(n_{\mathrm{L}})_{1}}\bigg]^{2}\,|F_{1}(x,y)|^{2}.\right.
\end{align}
This quantity only depends on \maths{v_{1}\,t-z} since \maths{(\Delta n_{\mathrm{L}})'(\mathbf{r},\omega_{1})\propto|A_{2}(z,t)|^{2}} and since \maths{A_{2}(z,t)} is assumed to propagate according to Eq.~\eqref{Eq:TwoCoreFiber}.

Upon quantization, we use Eq.~\eqref{Eq:FluxPoyntingVectorFiberClassical} with \maths{|A_{1}(z,t)|^{2}} replaced with \maths{\hat{A}_{1}^{\dag}(z,t)\,\hat{A}_{1}^{\vphantom{\dag}}(z,t)}. This yields Eq.~\eqref{Eq:FluxPoyntingVectorFiber} in the optical fiber (\maths{0<z<L}) and Eq.~\eqref{Eq:FluxPoyntingVectorFreeSpace} in free space (\maths{z<0} or \maths{z>L}).


\begin{thebibliography}{999}
\bibitem{NoteCoordinates}
Of course, it would have been more natural to define the free-space \maths{\tau} and \maths{\zeta} as \maths{\tau=z/c_{0}} and \maths{\zeta=c_{0}\,t-z}. This would have yielded Eq.~\eqref{Eq:MatterFieldFreeSpaceBis-a} with \maths{v_{1}/c_{0}=1}. However, using the definitions \eqref{Eq:TimeVariable} and \eqref{Eq:SpaceVariable} everywhere facilitates the matching of the fields at the \maths{z=0} and \maths{z=L} interfaces.
\end{thebibliography}
\end{document}